\documentclass[fleqn,usenatbib]{mnras}

\usepackage[T1]{fontenc}
\usepackage{ae,aecompl}

\usepackage{graphicx}	% Including figure files
\usepackage{amsmath}	% Advanced maths commands
\usepackage{amssymb}	% Extra maths symbols
\usepackage{longtable}
\usepackage{supertabular}
\usepackage{multirow}
\usepackage{multicol}

\usepackage{etoolbox}
\usepackage{newtxtext,newtxmath}

\title[EXO 1846$-$031 2019 outburst radio observations]{Radio observations of the Black Hole X-ray Binary EXO 1846$-$031 re-awakening from a 34-year slumber}

\author[D. R. A. Williams et al.]{D. R. A. Williams,$^{1,2}$\thanks{E-mail: david.williams$-$7@manchester.ac.uk}
S. E. Motta,$^{2,3}$
R. Fender,$^{2,4}$
J. C. A. Miller-Jones,$^{5}$
J. Neilsen,$^{6}$
J. R. Allison,$^{7,2}$
\newauthor{J. Bright,$^{2,8}$
I. Heywood,$^{2,9}$,
P. F. L. Jacob,$^{2}$
L. Rhodes,$^{2,10}$
E. Tremou,$^{11}$
P. Woudt,$^{4}$
J. van den Eijnden,$^{2}$}
\newauthor{F. Carotenuto,$^{2}$
D. A. Green,$^{12}$
D. Titterington,$^{12}$
A. J. van der Horst$^{13,14}$ and 
P. Saikia$^{15}$
}
\\
$^{1}$Jodrell Bank Centre for Astrophysics, School of Physics and Astronomy, The University of Manchester, Manchester, M13 9PL, UK\\
$^{2}$Department of Physics, University of Oxford, Denys Wilkinson Building, Keble Road, Oxford, OX1 3RH, UK \\
$^{3}$Istituto Nazionale di Astrofisica, Osservatorio Astronomico di Brera, via E. Bianchi 46, 23807 Merate (LC), Italy\\
$^{4}$Department of Astronomy, University of Cape Town, Private Bag X3, Rondebosch 7701, South Africa \\
$^{5}$International Centre for Radio Astronomy Research – Curtin University, Perth, Western Australia 6845, Australia\\
$^{6}$Department of Physics, Villanova University, Mendel Science Center, Villanova PA, 19085, USA\\
$^{7}$First Light Fusion Ltd., Unit 9/10 Oxford Industrial Park, Mead Road, Yarnton, Kidlington OX5 1QU, UK\\
$^{8}$Center for Interdisciplinary Exploration and Research in Astrophysics (CIERA) and Department of Physics and Astronomy, Northwestern University, \\
Evanston, IL 60208, USA\\
$^{9}$Department of Physics and Electronics, Rhodes University, PO Box 94, Grahamstown 6140, South Africa\\
$^{10}$Max-Planck-Institut f{\"u}r Radioastronomie, Auf dem H{\"u}gel 69, D-53121 Bonn, Germany\\
$^{11}$National Radio Astronomy Observatory, Socorro, NM 87801, USA\\
$^{12}$Astrophysics Group, Cavendish Laboratory, 19 J. J. Thomson Avenue, Cambridge CB3 0HE, UK\\
$^{13}$Department of Physics, The George Washington University, 725 21st Street NW, Washington, DC 20052\\
$^{14}$Astronomy, Physics, and Statistics Institute of Sciences (APSIS), 725 21st Street NW, Washington, DC 20052\\
$^{15}$Center for Astro, Particle and Planetary Physics (CAP$^3$), New York University Abu Dhabi, PO Box 129188, Abu Dhabi, UAE\\
}

\date{Accepted XXX. Received YYY; in original form ZZZ}

\pubyear{2022}

\begin{document}
\label{firstpage}
\pagerange{\pageref{firstpage}--\pageref{lastpage}}
\maketitle

\begin{abstract}
We present radio [1.3\,GHz MeerKAT, 4$-$8\,GHz Karl G. Jansky Very Large Array (VLA) and 15.5\,GHz Arcminute Microkelvin Imager Large Array (AMI-LA)] and X-ray (\textit{Swift} and \textit{MAXI}) data from the 2019 outburst of the candidate Black Hole X-ray Binary (BHXB) EXO 1846$-$031.
We compute a Hardness--Intensity diagram, which shows the characteristic q-shaped hysteresis of BHXBs in outburst. 
EXO 1846$-$031 was monitored weekly with MeerKAT and approximately daily with AMI-LA. The VLA observations provide sub-arcsecond-resolution images at key points in the outburst, showing moving radio components. 
The radio and X-ray light curves broadly follow each other, showing a peak on $\sim$MJD 58702, followed by a short decline before a second peak between $\sim$MJD 58731--58739.
We estimate the minimum energy of these radio flares from equipartition, calculating values of $E_{\rm min} \sim$ 4$\times$10$^{41}$ and 5$\times$10$^{42}$ erg, respectively. 
The exact date of the return to `quiescence' is missed in the X-ray and radio observations, but we suggest that it likely occurred between MJD 58887 and 58905. 
From the \textit{Swift} X-ray flux on MJD 58905 and assuming the soft-to-hard transition happened at 0.3--3 per cent Eddington, we calculate a distance range of 2.4--7.5\,kpc.
We computed the radio:X-ray plane for EXO 1846$-$031 in the `hard' state, showing that it is most likely a `radio-quiet' BH, preferentially at 4.5\,kpc. 
Using this distance and a jet inclination angle of $\theta$=73$^{\circ}$, the VLA data place limits on the intrinsic jet speed of $\beta_{\rm int} = 0.29c$, indicating sub-luminal jet motion.

\end{abstract}


%%%%%%%%%%%%%%%%%%%%%%%%%%%%%%%%%%%%%%%%%%%%%%%%%%

%%%%%%%%%%%%%%%%% BODY OF PAPER %%%%%%%%%%%%%%%%%%

\section{Introduction}
\label{sec:introduction}

% Something about archival X-ray transients and old x-ray monitors
X-ray binaries (XBs) are binary systems which contain a compact object, such as a neutron star (NS) or black hole (BH) where matter is transferred from a secondary, non-degenerate donor star, via an accretion disc around the compact object. The majority of BHXBs spend most of their lives in a  `\textit{quiescent}' state, accreting at low rates and undetectable in the X-rays for years to decades. 
BHXBs are commonly discovered when they enter an `\textit{outburst}'; their multi-band flux increases by many orders of magnitude as the accretion rate increases. 
Usually, after weeks to months, the BHXB will gradually transition from the `hard' power-law ($\Gamma \sim$ 1.5) dominated state into a `soft' state \citep[see,][for full descriptions of different X-ray spectral states]{RemillardMcClintock2006,BelloniMotta}, whereby the X-ray spectrum gradually becomes dominated by blackbody disc emission and the X-ray flux begins to decrease, through `intermediate' states \citep[see e.g.][]{HomanBelloni2005}. The intermediate states are defined by their X-ray timing and spectral characteristics. They can be rapid, sometimes lasting only hours \citep[][]{Belloni2006}. %
After weeks to months in the `soft' state, the BHXB moves back to the `hard' state at 0.3--3 per cent of the Eddington luminosity \citep[][]{Maccarone2003,Kalemci2013,VahdatMotlagh2019}, before fading back into quiescence. This hysteresis is best demonstrated on a `Hardness--Intensity Diagram' (HID), whereby the X-ray flux (or luminosity) is plotted against the X-ray spectral hardness \citep[e.g.][]{FenderBelloniGallo}. 

% Segue into more recent findings in the x-rays and radio
The different X-ray states are closely associated with radio properties of BHXBs  \citep{Corbel2000,FenderBelloni,FenderBelloniGallo}. During the `hard' state (and quiescence), a flat-spectrum ($S \propto \nu^\alpha$, with spectral index $\alpha \sim
0$), steady jet persists and is observed \citep[e.g. Cygnus X-1,][]{Stirling2001}. When a BHXB transitions out of the `hard' state, the steady jet is quenched by at least 2 orders of magnitude \citep[][\citealt{Russell2019}]{Coriat2011,Russell2011}. Transient ejecta can be emitted from the BHXB as the source passes through the `intermediate' states \citep[][]{Miller-Jones2012,Russell2019,Wood2021}, which may be associated with bright radio flares as a signature of the ejection \citep[][]{Corbel2004,FenderBelloniGallo}. These radio ejecta can be tracked as they move away from the BHXB and interact with the interstellar medium \citep[][]{MirabelRodriguez94,Corbel2002,Corbel2005,Bright2020,Carotenuto2021b}. Transient radio ejecta typically show radio spectral indices which steepen ($\alpha <$ 0) over time as the component expands and becomes optically-thin. 
If the distance is known, observations of BHXB radio jets can enable the calculation of fundamental properties of the jet and BH system 
\citep[][]{MirabelRodriguez94,Fender2006}. The steady jet only reappears when the source transitions back into the `hard' state and into quiescence. 

%disc jet coupling + R/X plane
During the X-ray `hard' state, the radio and X-ray behaviour of BHXBs is correlated and has been studied in depth for many sources using quasi-simultaneous observations \citep[e.g.][]{GalloFenderPooley,Corbel2003,Coriat2011,Corbel2013} and is known as the radio:X-ray plane. 
It was thought that all XBs followed a relation in the form of $L_{\rm Radio} \propto L_{\rm X-ray}^{0.6}$, based upon early observations of GX 339$-$4 \citep{Hannikainen1998,GalloFenderPooley,Corbel2003,Corbel2013}. This relation extends down to very low luminosities, i.e. into quiescence \citep{Corbel2003,Corbel2013,Plotkin2017,Tremou2020} and has been observed in other sources such as %A0620$-$00 \citep[][]{Gallo2006} 
V404 Cygni \citep[e.g.][]{Corbel2008} and XTE J1118+480 \citep[][]{Gallo2014}. 
However, further observations revealed the presence of another population of BHXBs which are less radio luminous than this relation, a so-called `radio-quiet' branch, which followed $L_{\rm R} \propto L_{\rm X}^{1.4}$, such as H~1743$-$322 \citep[e.g.][]{Coriat2011,Williams1743}. In some of these `radio-quiet' objects, they are then found to re-join the `radio-loud' branch when they go back into quiescence \citep{Coriat2011,Carotenuto2021}. The underlying cause of the split tracks \citep[see][for a clustering analysis into the statistical robustness of this split]{Gallo2012,Gallo2014,Gallo2018} for BHXBs is not known, but it may be due to differences in the radiative efficiency of the accretion flow \citep{Coriat2011,Koljonen2019}, an inclination effect of the source \citep[][]{MottaCasellaFender}, differences in the accretion disc contribution \citep[][]{Meyer2014} or changes in the magnetic field \citep[][]{Casella2009}. For the purposes of this paper, we will refer to the original $L_{\rm R} \propto L_{\rm X}^{0.6}$ correlation sources as `radio-loud' objects, and those that diverge onto the $L_{\rm R} \propto L_{\rm X}^{1.4}$ track as `radio-quiet' sources.

%\subsection{EXO 1846$-$031}
In this work, we investigate the 2019 outburst of the historical X-ray transient EXO 1846$-$031. EXO 1846$-$031 was first discovered as an `ultra-soft X-ray transient' by the \textit{EXOSAT} satellite during slewing manoeuvres on 1985 April 3 \citep[][]{Parmar1985}. Subsequent targeted \textit{EXOSAT} observations, optical and radio observations were performed, the results of which we summarise below: 
i) the X-ray properties indicated EXO 1846$-$031 was a BH candidate \citep{Parmar1993}; ii) the EXO 1846$-$031 field is crowded in the X-rays \citep{Parmar1993} and heavily confused in the radio \citep{NelsonSpencer}; iii) there is large absorption along the line-of-sight, $N_{\rm H} \sim$ 3.5$\times$10$^{22}$ atoms cm$^{-2}$ \citep{Parmar1993}; iv) no optical counterpart was found to a limiting magnitude of 21.5 mag \citep{WenzelOptical,Gottwald,ZhangOptical}; v) the distance is crudely estimated to be $\sim$7\,kpc, from scaling the first \textit{EXOSAT} observation peak flux to a luminosity of 10$^{38}$ erg s$^{-1}$ \citep{Parmar1993}.

Given the large number of unknown parameters of EXO 1846$-$031, the study of further outbursts is key for constraining the system properties. Despite reports of a possible outburst in 1994 \citep{ZhangOptical,Inoue1994}, no new bona-fide outbursts of EXO 1846$-$031 had been reported until 2019 July 31 \citep{NegoroATel}. On MJD 58687.864 (2019 July 23 20:44 UTC), the nova alert system on-board the \textit{MAXI/GSC} satellite had detected an increase in X-ray flux from the region around a nearby source IGR J18483$-$0311 \citep{NegoroATel}, but it was only realised a week later on MJD 58694.878 (2019 July 30 21:05 UTC) that the source of the flux increase was instead EXO 1846$-$031 re-awakening from a $\sim$34 year slumber. Following an increase in the 2--4 and 4--10\,keV fluxes seen by \textit{MAXI/GSC}, EXO 1846$-$031 was confirmed to be in the `hard' state and in outburst \citep{NegoroATel}. A further observation with the X-ray telescope (XRT) instrument onboard the Neil Gehrels \textit{Swift} Observatory gave an improved localisation \citep[$18^{\rm h}49^{\rm m}16\fs99~$-$03^{\circ}03\arcmin55\farcs4$, with a 90 per cent confidence error radius of 2~arcsec;][]{MereminskiyATel}. 

Analysis of the 2019 outburst of EXO 1846$-$031 has already been performed, showing: i) Type-C QPOs in \textit{NICER} \citep{BultATel} and \textit{Insight-HXMT} \citep{Liu2020} data; ii) a characteristic HID to that of a BHXB \citep{Liu2020}; 
iii) an ionised disc wind with a velocity up to $0.06c$ in the \textit{Insight-HXMT} data \citep{Wang2021}; iv) \citet{Draghis2020} argue that EXO 1846$-$031 is a BH with a maximal spin parameter at disc inclination of $\theta$=73$^{\circ}$, though this inclination is disputed by \citet{Wang2021} who use the same data and prefer $\theta$=40$^{\circ}$; iv) a compact radio source early in the outburst favoured a BH primary, as EXO 1846$-$031 would need to be closer than 3.7\,kpc to be consistent with the brightest detections of neutron stars \citep{Miller-JonesATel}. All of the literature arguments point toward a BH nature for the compact object in EXO 1846$-$031, though a NS cannot be ruled out as we still lack a dynamical mass measurement.

Following the multi-wavelength reports of renewed activity of EXO 1846$-$031, we triggered observations \citep[][]{WilliamsEXO} with the MeerKAT telescope as part of the  ThunderKAT\footnote{\textbf{T}he \textbf{hun}t for \textbf{d}ynamic and \textbf{e}xplosive \textbf{r}adio transients with Meer\textbf{KAT}: \url{http://www.thunderkat.uct.ac.za/}} Large Survey Project \citep{ThunderKAT2017}, as well as observations with the Arcminute Microkelvin Imager Large Array \citep[AMI-LA,][]{AMILA,AMILA2} as part of a long-term program to follow up astrophysical radio transients (PI: Fender). We also acquired Karl G. Jansky Very Large Array (VLA) observations (PIs: Miller-Jones, Neilsen) throughout the outburst. Alongside these observations, we triggered an X-ray monitoring program with \textit{Swift} to follow the new outburst in the X-ray waveband. This paper is structured as follows: in Section~2, we describe our observations and data reduction, in Section~3, we present our results and discuss their implications in Section~4, and finally we summarise our conclusions in Section~5.

\section{Observations and Data Reduction}

\subsection{Radio data of the 2019 outburst}
\label{sec:radioobs}

\subsubsection{Karl G. Jansky VLA data}

Observations of EXO 1846$-$031 were performed with the Karl G. Jansky Very Large Array (VLA). The first observation was obtained on MJD 58696.199 and showed integrated flux densities of 2.54$\pm$0.03\,mJy and 2.42$\pm$0.03\,mJy at 5.25 and 7.45\,GHz, respectively \citep[][]{Miller-JonesATel}. A subsequent observation (PI: Miller-Jones, 19A-217) was made two days later (MJD 58698.107). Both of these observations used the same 8-bit correlator setup: 2$\times$ basebands centred at 5.25 and 7.45\,GHz, each with eight spectral windows and a bandwidth of $\sim$1\,GHz. These observations used 3C286 as the flux calibrator and included three minutes of on-source time, as well as J1832$-$1035 as the phase calibrator. Four further observations (PI: Neilsen, SK0577) were obtained on MJD 58709.330, 58723.147, 58746.082 and 58776.012. All four observations were also performed in C band (6\,GHz), but in slightly different correlator configurations to the 19A-217 observations. The first observation (MJD 58709.330) used two 8-bit sub-bands both of $\sim$1\,GHz in bandwidth and eight spectral windows, centred at 4.7 and 7.4\,GHz, used 3C48 as the flux calibrator and included eighty-five minutes of on source time. The other three observations used two 3-bit basebands centred at 4.9 and 7\,GHz, each with a bandwidth of $\sim$2\,GHz, 16 spectral windows in each baseband and included seventy-five minutes of on-source time with 3C286 as the flux calibrator and J1832$-$1035 as the phase calibrator. All six observations were made whilst the VLA was in A-configuration with baselines up to 36~km, yielding $<$0.5-arcsec angular resolution.

We downloaded the data from the VLA Archive\footnote{\url{https://data.nrao.edu/portal/}} and performed initial calibration procedures using the VLA \textsc{CASA} pipeline version 5.6.3 \citep{CASA}. After manual excision of some radio frequency interference (RFI), removal of bad antennas, we re-ran the pipeline. After final inspection of the data, we split the target field from the data and averaged the visibilities to 10s to speed up the imaging process. 
We used \texttt{tclean} in \textsc{CASA} to make initial images with an image size of 300$\times$300 pixels, a Nyquist sampled cell size between 0.04 and 0.05 arcsec, depending on the observation and Briggs weighting with a robust parameter of 0.5. A further image with a robust parameter of $-$1 was made to highlight compact structures.  
We also performed additional imaging runs in both sub-bands for all observations, so that a two-point radio spectral index could be estimated in the VLA data (see Section~\ref{sec:specindex}). We used \textsc{imfit} to extract flux densities and fold in a conservative 10 per cent flux calibration error\footnote{3C48 has been varying since 2018, see: \url{https://science.nrao.edu/facilities/vla/docs/manuals/oss/performance/fdscale}} to take account of the varying flux calibrator used. Positional uncertainties of the VLA are estimated to be 10 per cent the synthesized beam, which we fold into the statistical fit errors in quadrature. Positions and source parameters are reported in Table~\ref{tab:VLAdata}. The final observation shows two clear components, which are reported separately in Table~\ref{tab:VLAdata}.

\begin{table*}
    \centering
    \caption{Radio data for EXO 1846$-$031 for VLA. In all cases the top row for a given date is the result for the lower frequency band and the bottom row is the result for the higher frequency band. Where a single value is given, this corresponds to the full bandwidth result. The columns in the table are as follows: (1) the start date and time in UTC for the observation; (2) the central MJD for the on source scan; (3) the central frequency of the sub-bands; (4) the brightness for each sub-band obtained from \textsc{CASA} fitting procedures described in Section~\ref{sec:radioobs} in mJy beam$^{-1}$; (5) integrated flux density in mJy for the full bandwidth of the given observation; (6) the rms noise level obtained in a region near to the source for the full band-width in $\upmu$Jy beam$^{-1}$ and used as the rms level in Fig.~\ref{fig:radioejecta}; (7) the synthesized beam size in arcsec$^2$ for the full bandwidth image; (8) the position angle in degrees east of north for the full bandwidth image; 
    (9) Right Ascension (top row) and associated error (bottom row) obtained from the \textsc{imfit} procedure; (10) Declination (top row) and associated error (bottom row) obtained from the \textsc{imfit} procedure. 
	The flux densities listed in this table include a 10 per cent calibration error added in quadrature (see Section~\ref{sec:radioobs}). The positional errors represent 10 per cent of the synthesized beam added in quadrature to the statistical fitting error.}
    \begin{tabular}{c c c c c c c c c c }
    \hline
         Start Date &  Central & Freq. & Peak Brightness & Integrated & RMS Noise  & Beam & P.A. & Right Asc. & Declination \\
         + Time & MJD & (GHz) & (mJy beam$^{-1}$) & Flux (mJy) & ($\upmu$Jy beam$^{-1}$). & arcsec$^{2}$ & $^\circ$ & + err. & + err.\\
         (1) & (2) & (3) & (4) & (5) & (6) & (7) & (8) & (9) & (10) \\
         \hline
         \multicolumn{10}{c}{\textit{Central component}}\\
         \hline
         2019-08-01 & \multirow{2}{*}{58696.199}& 5.25 & 2.5$\pm$0.3 & \multirow{2}{*}{2.5$\pm$0.3} & \multirow{2}{*}{50} & \multirow{2}{*}{0.35$\times$0.27} & \multirow{2}{*}{$-$3.1} & 18:49:17.047 & $-$03:03:55.24 \\
         04:45:55 & & 7.45 & 2.4$\pm$0.3 & &  & & &  %$\pm$0\fs0001 &
         $\pm$0\fs003 &
         %$\pm$0\farcs004 
         $\pm$0\farcs04 \\
         \hline
         2019-Aug-03 & \multirow{2}{*}{58698.107}& 5.25 & 5.9$\pm$0.6 & \multirow{2}{*}{6.0$\pm$0.6} & \multirow{2}{*}{35} & \multirow{2}{*}{0.43$\times$0.26} & \multirow{2}{*}{$-$36.0} & 18:49:17.047 & $-$03:03:55.25 \\
         02:34:21 & & 7.45 & 6.1$\pm$0.6 & &  & & &  %$\pm$0\fs00005 &
         $\pm$0\fs003 &%$\pm$0\farcs0009
         $\pm$0\farcs04\\         
         \hline
         2019-Aug-14 & \multirow{2}{*}{58709.330}& 4.70 & 2.9$\pm$0.3 & \multirow{2}{*}{2.7$\pm$0.3} & \multirow{2}{*}{42} & \multirow{2}{*}{0.47$\times$0.27} & \multirow{2}{*}{60.5} & 18:49:17.056 & $-$03:03:55.27 \\
         07:55:48 & & 7.40 & 1.5$\pm$0.2 & &  & & &  %$\pm$0\fs0006 &
         $\pm$0\fs003 & %$\pm$0\farcs007
         $\pm$0\farcs05 \\            
         \hline
         2019-Aug-28 & \multirow{2}{*}{58723.147}& 5.00 & 1.14$\pm$0.10 & \multirow{2}{*}{1.11$\pm$0.11} & \multirow{2}{*}{10} & \multirow{2}{*}{0.33$\times$0.25} & \multirow{2}{*}{2.0} & 18:49:17.048 & $-$03:03:55.26 \\
         03:31:20 & & 7.00 & 0.95$\pm$0.10 & &  & & &  %$\pm$0\fs0001 &
         $\pm$0\fs002 &%$\pm$0\farcs002 \\
         $\pm$0\farcs03 \\ 
         \hline
         2019-Sep-20 & \multirow{2}{*}{58746.082}& 5.00 & 11.8$\pm$1.2 & \multirow{2}{*}{15.6$\pm$1.6} & \multirow{2}{*}{42} & \multirow{2}{*}{0.34$\times$0.28} & \multirow{2}{*}{119.9} & 18:49:17.048 & $-$03:03:55.28 \\
         01:57:54 & & 7.00 & 9.5$\pm$1.0 & &  & & &  %$\pm$0\fs0001 &%
         $\pm$0\fs002 &%$\pm$0\farcs002 \\
         $\pm$0\farcs03 \\
         \hline
         2019-Oct-20 & \multirow{2}{*}{58776.012}& 5.00 & 0.23$\pm$0.02 & \multirow{2}{*}{0.31$\pm$0.03} & \multirow{2}{*}{8} & \multirow{2}{*}{0.37$\times$0.26} & \multirow{2}{*}{$-$49.0} & 18:49:17.045 & $-$03:03:55.27 \\
         00:17:54 & & 7.00 & 0.15$\pm$0.02 & &  & & &  %\pm$0\fs0003 &%
         $\pm$0\fs004 &%$\pm$0\farcs002 \\
         $\pm$0\farcs02 \\
         \hline
         \multicolumn{10}{c}{\textit{Second component}}\\
         \hline
         2019-Oct-20 & \multirow{2}{*}{58776.012}& 5.00 & 0.16$\pm$0.02 & \multirow{2}{*}{0.15 $\pm$0.02} & \multirow{2}{*}{8} & \multirow{2}{*}{0.37$\times$0.26} & \multirow{2}{*}{$-$49.0} & 18:49:17.103 & $-$03:03:55.37 \\
         00:17:54 & & 7.00 & 0.10$\pm$0.02 & &  & & &  %$\pm$0\fs0003 &
         $\pm$0\fs004 &%$\pm$0\farcs006 \\
         $\pm$0\farcs02 \\

    \hline    
    \end{tabular}
    \label{tab:VLAdata}
\end{table*}

\subsubsection{MeerKAT data}

ThunderKAT observations of EXO 1846$-$031 began on 2019 August 04 (MJD 58699) and continued with approximately weekly cadence until 2020 January 03 (MJD 58851). An additional two observations were performed on 2020 February 21 (MJD 58900) and 2020 April 10 (MJD 58949). In total we obtained 25 epochs, all of which are listed in Table~\ref{tab:radiodataMeerKAT}, along with key observation properties. All MeerKAT observations were performed at a central frequency of 1.284\,GHz, with a bandwidth of 856\,MHz, an integration time of 8~s, and consisted of one 15 minute scan of EXO 1846$-$031, bracketed by 2 minute scans of the secondary calibrator (PKS J1911$-$2006). Each observation included a scan of PKS J1939$-$6342 as the primary calibrator, to set the band pass and flux scale. Depending on the elevation of EXO 1846$-$031, each observation resulted in a angular resolution of 6--12~arcsec. 

The data were calibrated using \textsc{CASA} Version 5.6.2, within the \texttt{OxKAT} reduction package\footnote{\url{https://github.com/IanHeywood/oxkat}} \citep{OxKAT}. We made \textit{a priori} excision of RFI\footnote{A combination of manual flags from known areas of RFI in the MeerKAT band plus automated identification and removal of RFI were used.} and then split the data into eight spectral windows. The delay, band pass and gain calibrations were calculated on the primary calibrator field and then applied to the secondary calibrator field. We accounted for the spectral index of the secondary calibrator and applied the solutions onto the target fields. Finally, we performed some additional flagging of RFI on the target fields before imaging. Imaging of the target field was performed using version 2.9.0 of \textsc{wsclean} \citep{wsclean,wsclean2}, using a threshold mask generated from the data. Due to a large amount of bright diffuse structure in the $\sim$2$^{\circ}$ square target field, the median rms noise threshold achieved was  $\sim 70$~$\upmu$Jy beam$^{-1}$. For imaging, we used a pixel size of 1.1-arcsec and used a Briggs weighting with a robust parameter of $-0.85$. A single round of phase-only self-calibration was performed with a 64~s solution interval. We used the \textsc{CASA} task \textsc{imfit} to extract source positions and fluxes, folding a 10 per cent flux calibration uncertainty due to underlying systematic variations in the data \citep[][]{Driessen2020} in quadrature into our flux density values, which we report in Table~\ref{tab:radiodataMeerKAT}. We also made images in each sub-band and performed the same fitting analysis for each epoch, which allowed us to obtain in-band spectral indices for EXO 1846$-$031 (see Section~\ref{sec:specindex}). 

\begin{table}
 \caption{Radio data for EXO 1846$-$031 for MeerKAT. The columns in the table are as follows: (1) the central MJD for the on source scan; (2) the peak brightness obtained from \textsc{CASA} fitting procedures described in Section~\ref{sec:radioobs} in mJy~beam$^{-1}$; (3) the integrated flux density in mJy; (4) the rms noise level obtained in a region near to the source; (5) the synthesized beam in units of arcseconds squared and the position angle, in units of degrees east of north.
 Upper limits are denoted with `<' and are given at the 3$\times$rms-noise level. }
	\centering
	\begin{tabular}{ c c c c c  }
    \hline
    Central & Peak  & Int. Flux & RMS  & Beam Size / P.A. \\       
    MJD    & Brightness    & Density & noise  & arcsec$^2$ / $^{\circ}$  \\
     & (mJy & (mJy) & $\rm \mu$Jy \\
     & ~beam$^{-1}$) & & ~beam$^{-1}$ \\ 
    (1)          & (2)     & (3)                   & (4)           & (5)          \\                          \\
    \hline
   58699.839 & 6.7$\pm$0.7 & 6.9$\pm$0.7 & 121 & 8.08$\times$5.33 / $-35.32$\\%& $0.07^{+0.19}_{-0.18}$ \\ 
   58705.803 & 30.4$\pm$3.1 & 30.8$\pm$3.1 & 91 & 8.07$\times$5.56 / $-38.39$\\%& $-0.48^{+0.12}_{-0.12}$ \\ 
   58711.904 & 12.3$\pm$1.2 & 12.4$\pm$1.3 & 69 & 7.11$\times$7.08 / $-10.3$\\%& $-0.69^{+0.13}_{-0.12}$ \\ 
   58718.695 & 5.4$\pm$0.5 & 5.4$\pm$0.6 & 69 & 8.46$\times$6.69 / $-61.19$\\%& $-0.35^{+0.17}_{--0.17}$ \\ 
   58726.779 & 18.1$\pm$1.8 & 18.2$\pm$1.8 & 67 & 7.19$\times$5.38 / $-26.78$\\%& $-0.3^{+0.12}_{--0.13}$ \\ 
   58733.692 & 34.8$\pm$3.5 & 34.9$\pm$3.5 & 84 & 8.34$\times$5.82 / $-44.78$\\%& $-0.25^{+0.11}_{-0.11}$ \\ 
   58740.782 & 32.3$\pm$3.2 & 32.5$\pm$3.3 & 67 & 6.73$\times$5.80 / $-28.2$\\%& $-0.2^{+0.12}_{-0.12}$\\ 
   58747.654 & 21.4$\pm$2.1 & 21.6$\pm$2.2 & 80 & 8.27$\times$5.69 / $-45.02$\\%& $-0.32^{+0.11}_{-0.12}$\\ 
   58755.666 & 8.4$\pm$0.9 & 8.6$\pm$0.9 & 98 & 7.61$\times$5.37 / $-34.91$\\%& $-0.35^{+0.14}_{-0.14}$\\ 
   58762.684 & 3.7$\pm$0.4 & 3.7$\pm$0.4 & 73 & 7.89$\times$5.42 / $-34.9$\\%& $-0.52^{+0.18}_{-0.19}$\\
   58768.762 & 2.0$\pm$0.2 & 2.1$\pm$0.2 & 102 & 8.33$\times$6.37 / $-71.29$\\%& $-1.19^{+0.26}_{-0.27}$\\ 
   58775.646 & 1.5$\pm$0.2 & 1.5$\pm$0.2 & 89 & 7.47$\times$5.55 / $-32.26$\\%& $-1.58^{+0.44}_{-0.39}$\\ 
   58782.639 & 0.9$\pm$0.1 & 1.3$\pm$0.2 & 79 & 6.94$\times$5.42 / $-22.03$\\%& $-0.69^{+0.81}_{-0.82}$\\ 
   58788.720 & 0.5$\pm$0.1 & 1.0$\pm$0.3 & 104 & 7.62$\times$6.64 / $-38.73$\\%& $-0.82^{+0.9}_{-0.87}$\\ 
   58797.445 & 0.4$\pm$0.1 & 0.7$\pm$0.2 & 64 & 9.01$\times$6.85 / $75.05$\\%& $-1.63^{+1.17}_{-1.15}$\\  
   58804.439 & 0.4$\pm$0.1 & 0.4$\pm$0.1 & 73 & 8.24$\times$6.98 / $83.76$\\%& $-0.69^{+0.95}_{-0.99}$\\ 
   58811.381 & 0.5$\pm$0.1 & 0.5$\pm$0.1 & 88 & 12.04$\times$6.75 / $75.9$\\%& $-1.98^{+1.06}_{-1.06}$\\ 
   58817.486 & - & $<$0.2 & 58 & 7.72$\times$5.41 / $-38.89$\\%& - \\ 
   58824.359 & 0.7$\pm$0.1 & 1.0$\pm$0.2 & 67 & 10.37$\times$6.52 / $64$\\%& $-1.55^{+0.93}_{-0.87}$\\ 
   58832.427 & 0.7$\pm$0.1 & 0.8$\pm$0.1 & 64 & 8.32$\times$5.83 / $-42.3$\\%& $-1.53^{+0.91}_{-0.88}$\\                       
   58837.469 & - & $<$0.2 & 75 & 7.42$\times$5.51 / $-27.84$\\%& - \\ 
   58845.312 & - & $<$0.2 & 68 & 10.36$\times$6.57 / $57.48$\\%& - \\ 
   58851.438 & - & $<$0.2 & 77 & 6.73$\times$5.88 / $-18.56$\\%& - \\ 
   58900.274 & - & $<$0.2 & 69 & 7.50$\times$5.50 / $-34.15$\\%& - \\ 
   58949.314 & - & $<$0.2 & 78 & 9.62$\times$6.41 / $-47.41$\\%& - \\    

    \hline

\end{tabular}
	\label{tab:radiodataMeerKAT}
\end{table}

\subsubsection{AMI-LA data}

AMI-LA observations began on 2019 August 06 (MJD 58701) and continued approximately daily until 2019 October 07 (MJD 58763), occasionally missing some days due to operational issues such as high winds. The early observations were often less than two hours in duration. After the initial pseudo-daily monitoring, EXO 1846$-$031 was observed approximately weekly, often within 1$-$2 days of the MeerKAT observations, to obtain bichromatic quasi-simultaneous radio data. The longest observation during this phase was four hours. Observations ceased on 2020 March 14 (MJD 58922) due to the instrument shutdown caused by the  COVID-19 pandemic. 

All AMI-LA observations were performed at a central frequency of 15.5\,GHz with a bandwidth of 5\,GHz \citep{AMILA, AMILA2}. The maximum baseline length of AMI-LA is $\sim$110~m, leading to angular resolutions of approximately 30 arcsec, though the southern nature of EXO 1846$-$031 leads to a significantly elongated north-south major axis (up to 50 arcsec). All observations used a standard phase referencing technique, with 10-min scans on the target source interleaved with 100~s scans of the phase calibrator source NVSS J185146+003532. Daily observations of the flux calibrator 3C286 were used to bootstrap the flux density scale. We estimated that the flux calibration error was 8 per cent from the flux densities of the phase calibrator, which we fold in quadrature into our fluxes. The AMI-LA campaign is summarised in Table~\ref{tab:radiodataAMI}. Two observations, obtained on 2019 September 30 and 2020 February 27 are included, but due to heavy rain and snow respectively, the instrument sensitivity is severely affected. 

To calibrate the data, we used a custom reduction package \texttt{REDUCE$\_$DC} \citep{Davies2009,Perrott2013,Perrott2015} and then proceeded to image the data using the Common Astronomical Software Applications \textsc{CASA} software \citep{CASA} task \texttt{clean}, restricting the \textit{uv}-range to $\ga$1400$\lambda$ to remove the contribution of a diffuse source near to the phase centre. The \textsc{CASA} task \texttt{imfit} yielded flux densities, which are reported in Table~\ref{tab:radiodataAMI}.

\begin{table}
    \centering
    \caption{Radio data for EXO 1846$-$031 for AMI-LA: (1) shows the central MJD of the observation which is used in the figures in this work; (2) shows the total time on source in hours; (3) reports the peak brightness in mJy beam$^{-1}$, with the 8 per cent calibration added in quadrature (see Section~\ref{sec:radioobs}); (4) shows the rms noise level in an off-source region near the phase centre of the image. The observations marked with an `*' are severely affected by adverse weather conditions. The `\textdagger'  indicates an observation where significant flagging of one half of the band was performed due to poor rain gauge data, making it impossible to produce a spectral index across the band, as shown in Table~\ref{tab:spec_index}. }
    \begin{tabular}{ c c c c }
    \hline Central &  Obs.   &   Peak Brightness       & RMS noise   \\
MJD   & dur. (h) & (mJy beam$^{-1}$) & (mJy beam$^{-1}$) \\
(1) & (2) & (3) & (4) \\
\hline
58701.841  & 0.4 & 5.9$\pm$0.5 & 0.86\textdagger \\
  58702.876  & 1.5 & 17.4$\pm$1.4 & 0.33 \\
  58703.905  & 2 & 5.5$\pm$0.4 & 0.22 \\
  58706.906  & 2 & 2.5$\pm$0.3 & 0.26\textdagger \\
  58707.857  & 1 & 1.5$\pm$0.2 & 0.29 \\
  58708.867  & 1 & 2.3$\pm$0.2 & 0.25 \\
  58709.876  & 1 & 4.0$\pm$0.3 & 0.18 \\
  58710.925  & 2 & 2.8$\pm$0.3 & 0.43 \\
  58711.824  & 1 & 3.0$\pm$0.3 & 0.35 \\
  58712.931  & 0.3 & 2.3$\pm$0.3 & 0.43 \\
  58713.856  & 2 & 1.8$\pm$0.1 & 0.23\textdagger \\
  58714.907  & 1 & 1.7$\pm$0.1 & 0.26 \\
  58715.865  & 2 & 1.3$\pm$0.2 & 0.21 \\
  58717.828  & 2 & 5.3$\pm$0.4 & 0.24 \\
  58719.816  & 1 & <3.6        & 1.25 \\
  58724.811  & 2 & 4.7$\pm$0.4 & 0.27 \\
  58726.893  & 1 & 6.4$\pm$0.5 & 0.27  \\
  58728.841  & 2 & 18.5$\pm$1.5 & 0.14  \\
  58729.791  & 2 & 18.6$\pm$1.5 & 0.22  \\
  58730.811  & 2 & 22.9$\pm$1.8 & 0.17  \\
  58731.843  & 1 & 21.8$\pm$1.7 & 0.24  \\
  58733.869  & 0.4 & 15.3$\pm$1.2 & 0.31  \\
  58734.839  & 1 & 16.8$\pm$1.3 & 0.22  \\
  58735.887  & 0.4 & 18.0$\pm$1.5 & 0.45  \\
  58738.836  & 1 & 22.7$\pm$1.8 & 0.29  \\
  58739.730  & 0.4 & 11.8$\pm$1.0 & 0.55 \\
  58740.878  & 0.25 & 15.0$\pm$1.2 & 0.55 \\
  58741.800  & 0.4 & 13.5$\pm$1.1 & 0.71 \\
  58742.799  & 1 & 12.6$\pm$1.0 & 0.21 \\
  58744.791  & 1 & 11.9$\pm$1.0 & 0.31 \\
  58745.790  & 1 & 10.3$\pm$0.8 & 0.36 \\
  58747.785  & 1 & 10.2$\pm$0.8 & 0.35 \\
  58749.770  & 1 & 11.2$\pm$0.9 & 0.32 \\
  58750.811  & 0.6 & 9.5$\pm$0.8 & 0.44 \\
  58751.816  & 1 & 6.7$\pm$0.6 & 0.30 \\
  58752.794  & 1 & 4.3$\pm$0.4 & 0.22 \\
  58756.824*  & 0.2 & <7.1      & 2.35 \\
  58758.762  & 2 & 1.7$\pm$0.2 & 0.19 \\
  58759.729  & 2 & 1.3$\pm$0.1 & 0.21\textdagger \\
  58762.753  & 3.4 & 1.0$\pm$0.1 & 0.21 \\
  58763.738  & 4 & 1.5$\pm$0.1 & 0.14 \\
  58770.719  & 4 & <0.42 & 0.14 \\
  58777.699  & 4 & <0.54 & 0.18 \\
  58785.678  & 4 & <0.51 & 0.17 \\
  58787.662  & 3 & <0.42 & 0.14 \\
  58791.661  & 4 & <0.42 & 0.14 \\
  58796.608  & 2 & <0.57 & 0.19 \\
  58803.630  & 4 & <0.33 & 0.11 \\
  58810.611  & 4 & <0.51 & 0.17 \\
  58817.592  & 4 & <0.54 & 0.18 \\
  58824.573  & 4 & <0.42 & 0.14 \\
  58831.517  & 2.25 & <0.57 & 0.19 \\
  58838.534  & 4 & <0.48 & 0.16 \\
  \hline
  \end{tabular}
    \label{tab:radiodataAMI}
\end{table}

\begin{table}
    \centering
    \contcaption{ data from Table~\ref{tab:radiodataAMI}.}
  \begin{tabular}{ c c c c }
    \hline Central &  Obs.   &   Peak   Brightness      & RMS noise   \\
MJD   & dur. (h) &  (mJy beam$^{-1}$) & (mJy beam$^{-1}$) \\
(1) & (2) & (3) & (4)  \\
\hline
  58845.515  & 4 & <0.51 & 0.17 \\
  58852.496  & 4 & <0.39 & 0.13 \\
  58859.434  & 2 & <0.45 & 0.15 \\
  58866.460  & 4 & <0.51 & 0.17 \\
  58873.439  & 4 & <0.48 & 0.16 \\
  58881.417  & 4 & <0.84 & 0.28 \\
  58887.401  & 4 & <1.05 & 0.35 \\
  58899.320  & 1.35 & <0.93 & 0.31 \\
  58906.283*  & 0.8 & <2.40 & 0.80 \\
  58915.324  & 4 & <0.69 & 0.23 \\
  58922.305  & 4 & <0.63 & 0.21 \\
  \hline
    \end{tabular}
    \label{tab:radiodataAMI2}
\end{table}

\subsection{Radio Spectral Indices}

\label{sec:specindex}

The radio spectral index, $\alpha$, is a useful diagnostic for understanding the radio emission mechanism throughout the outburst (see Section~\ref{sec:introduction}). 
It is preferable to calculate a spectral index across a wider band, e.g. between AMI-LA (15.5\,GHz) and MeerKAT (1.3\,GHz). However, the fluxes clearly vary on a day-to-day timescale, so simultaneity (i.e. $<$1 day difference) of data from different arrays is necessary to obtain reliable measurements of the spectral index. This was only possible on six occasions with AMI-LA and MeerKAT detections and five others when AMI-LA could provide an upper limit. Similarly, we were able to obtain a VLA/AMI-LA radio spectral index on two occasions and one VLA/MeerKAT radio spectral index.
To supplement these radio spectral indices, we also calculated in-band spectra for MeerKAT, AMI-LA and VLA. All of these radio spectral indices are 
tabulated in Table~\ref{tab:spec_index}. For the MeerKAT data, we made an image for each of the eight spectral windows and fit a point source for EXO 1846$-$031 in each band for each epoch, folding in the flux density bootstrapping uncertainty. We fit a radio spectral index, taking into account upper limits in spectral windows where the source was undetected with the \textsc{linmix} package \citep{LINMIX}. The median value from the \textsc{linmix} fitting plus the 16th and 84th quartiles are reported as the spectral indices and errors. For the AMI-LA and VLA in-band spectra, we split the band into a higher and lower frequency band, and imaged each of these separately. We fit EXO 1846$-$031 in each band and fold in uncertainties in the flux density into the calculation of the spectral index \citep{EspinasseFender}. All of our radio light curves and spectral indices are shown in the first and second panels of Fig.~\ref{fig:lc}, respectively.

\begin{table*}
\caption{Table of \textit{quasi-simultaneous} spectral indices. Quasi-simultaneity refers to observations observed within 24 hours of one another, where there is no large day-to-day variability in the radio light curves (see Section~\ref{sec:specindex} for details). The table lists the following: (1) the MJD (see notes below); (2) the MeerKAT (1.28\,GHz) In-band spectral index; (3) the AMI-LA (15.5\,GHz) In-band spectral index; (4) the VLA ($\sim$5.5\,GHz) In-band spectral index; (5) the spectral index calculated from the AMI-LA and MeerKAT data; (6) the spectral index calculated from the VLA and AMI-LA; (7) the spectral index calculated from the VLA and MeerKAT data. 
The MJD date refers to one of the arrays listed as `M' for MeerKAT, `A' for AMI-LA and `V' for VLA. The first preference is to use the AMI-LA date, followed by the VLA date and finally the MeerKAT date. 
For MJD 58776.012, the VLA detection clearly shows two distinct sources (see Section~\ref{sec:JetSpeed}). For this date, we provide a MeerKAT In-band spectral index which includes both sources (core is unmarked, jet component marked with a `*'), as well as VLA spectra for each component separately. For the VLA/MeerKAT spectral index, we use the integrated flux densities of both components in the VLA data as the flux density value to compare to the unresolved point source flux density values in the MeerKAT data. }
    \centering
    \begin{tabular}{c c c c c c c }
    \hline
        MJD & MeerKAT & AMI-LA & VLA     & AMI-LA/ & VLA/   & VLA/\\
            & In-Band & In-band& In-Band & MeerKAT & AMI-LA & MeerKAT\\
        (1) & (2) & (3) & (4) & (5) & (6) & (7) \\
    \hline
        58696.199$^{\rm V}$ & - & - & $-$0.1$\pm$0.4 & -  & - & -\\ 
        58698.107$^{\rm V}$ & - & - & 0.1$\pm$0.4 & -  & - & -\\ 
        58699.839$^{\rm M}$ & 0.1$\pm$0.2 & - & - & - & - & -\\
        %58701.841 & AMI \textdagger\\
        58702.876$^{\rm A}$ & - & $-$0.7$\pm$0.7 & - & -  & - & -\\
        58703.905$^{\rm A}$ & - & $-$1.3$\pm$0.8 & - & -  & - & -\\
        58705.803$^{\rm M}$ & $-0.5\pm0.1$ & - & - & - & - & -\\
        %58706.906 & AMI \textdagger\\
        58707.857$^{\rm A}$ & - & $-$4.3$\pm$0.9 & -  & -  & - & -\\
        58708.867$^{\rm A}$ & - & $-0.3\pm0.9$ & $-1.5\pm0.3$  & -  & $-0.1\pm0.2$ & -\\
        58709.876$^{\rm A}$ & - & $-$0.2$\pm$0.8 & - & - & - & -\\
        58710.925$^{\rm A}$ & - & $-$1.4$\pm$0.8 & - & - & - & -\\
        58711.824$^{\rm A}$ & $-0.7\pm0.1$ & $-1\pm1$ & - & $-0.56\pm0.05$  & - & -\\ 
        58712.931$^{\rm A}$  & - & $-1.0\pm$0.9 & - & -  & - & -\\
        %58713.856 & AMI \textdagger\\
        58714.907$^{\rm A}$  & - & $-1.7\pm$1.3 & - & -  & - & -\\
        58715.865$^{\rm A}$  & - & $-2.6\pm$1.3 & - & -  & - & -\\
        58717.828$^{\rm A}$  & - & $-1.2\pm$0.7 & - & -  & - & -\\
        58718.695$^{\rm M}$ & $-0.4\pm0.2$ & - & - & - & - & -\\
        %58719.816 & AMI no det \\
        58723.147$^{\rm V}$ & - & - & $-$0.5$\pm$0.4 & - & - & -\\ 
        58724.811$^{\rm A}$  & - & $-1.7\pm$0.8  & - & - & - & -\\
        58726.893$^{\rm A}$ & $-$0.3$\pm$0.1 & $-0.1\pm$0.7 & - & $-$0.42$\pm$0.05  & - & -\\
        58728.841$^{\rm A}$  & - & $-0.4\pm$0.7  & - & - & - & -\\
        58729.791$^{\rm A}$  & - & $-0.7\pm$0.7  & - & - & - & -\\
        58730.811$^{\rm A}$  & - & $-0.4\pm$0.7  & - & - & - & -\\
        58731.843$^{\rm A}$  & - & $-0.4\pm$0.7  & - & - & - & -\\
        58733.869$^{\rm A}$ & $-0.3\pm0.1$ & $-0.3\pm$0.7 & - & $-0.33\pm0.05$ & - & - \\
        58734.839$^{\rm A}$  & - & $-$0.4$\pm$0.7 & - & - & - & -\\
        58735.887$^{\rm A}$  & - & $-$0.7$\pm$0.7 & - & - & - & -\\
        58738.836$^{\rm A}$  & - & $-$0.3$\pm$0.7 & - & - & - & -\\
        58739.730$^{\rm A}$  & - & $-$1.7$\pm$0.8 & - & - & - & -\\
        58740.878$^{\rm A}$ & $-0.2\pm0.1$ & $-0.5\pm0.7$ & - & $-0.31\pm0.05$ & - & - \\
        58741.800$^{\rm A}$ & - & 0.5$\pm$0.8  & - & - & - & -\\
        58742.799$^{\rm A}$ & - & $-$0.5$\pm$0.7  & - & - & - & -\\
        58744.791$^{\rm A}$ & - & $-$0.3$\pm$0.7  & - & - & - & -\\
        58745.790$^{\rm A}$ & - & $-$0.5$\pm$0.7  & $-$0.7$\pm$0.4 & - & $-$0.4$\pm$0.1 & -\\
        %58746.082$^{\rm V}$ & - & - &  & - \\ 
        58747.785$^{\rm A}$ & $-$0.3$\pm$0.1 & $-0.4\pm$0.7 & - & $-$0.30$\pm$0.05 & - & -\\
        58749.770$^{\rm A}$ & - & $-$1.8$\pm$0.7 & - & - & - & -\\
        58750.811$^{\rm A}$ & - & $-$1.0$\pm$0.7 & - & - & - & -\\
        58751.816$^{\rm A}$ & - & $-$1.0$\pm$0.8 & - & - & - & -\\
        58752.794$^{\rm A}$ & - & $-$0.2$\pm$0.8 & - & - & - & -\\
        58755.666$^{\rm M}$ & $-0.4\pm0.1$ & - & - & - & - & -\\
        %58756.824 & AMI no det \\
        58758.762$^{\rm A}$ & - & $-1.7\pm1.2$ & - & - & - & -\\
        %58759.729 & AMI \textdagger \\
        58762.753$^{\rm A}$ & $-0.5\pm0.2$ & $-0.5\pm1.2$ & - & $-0.49\pm0.06$ & - & -\\
        58763.738$^{\rm A}$ & - & $-3\pm1$ & - & - & - & -\\
        58768.762$^{\rm M}$ & $-1.2\pm0.3$ & - & - & - & - & -\\
        %58770.719 & AMI no det \\
        58776.012$^{\rm V}$ & \multirow{2}{*}{$-1.6\pm0.4$}& - & $-1.3\pm0.5$ & - & - & \multirow{2}{*}{$-0.8\pm0.1$}\\
        58776.012$^{\rm V}$* &  & - & $-$1.4$\pm$0.7 & -  & - & \\
        %58777.699 & AMI no det \\
        58782.639$^{\rm M}$ & $-$0.7$\pm$0.8& - & - & - & - & -\\
        %58785.678 & AMI no det \\
        %58787.662 & AMI no det \\
        58788.720$^{\rm M}$ & $-$0.8$\pm$0.9& - & - & - & - & -\\
        %58791.661 & AMI no det \\
        58796.608$^{\rm A}$ & $-$1.6$\pm$1.1& - & - & $<0.1$ & - & -\\ 
        %58797.445 
        58803.630$^{\rm A}$ & $-$0.7$\pm$1.0& - & - & $<-0.1$ & - & -\\
        %58804.439 & 
        58810.611$^{\rm A}$ & $-$2$\pm$1& - & - & $<0.1$ & - & -\\
        %58811.381 & 
        %58817.592 & AMI no det \\
        58824.573$^{\rm A}$ & $-$1.6$\pm$0.9& - & - & $<-0.2$ & - & -\\
        58831.517$^{\rm A}$ & $-$1.5$\pm$0.9& - & - & $<-0.1$ & - & -\\ 
        %58832.427 & 
    \hline
    \end{tabular}
    \label{tab:spec_index}
\end{table*}

\subsection{X-ray data of the 2019 outburst}
\label{sec:xrayobs}

\subsubsection{\textit{Swift}}\label{sec:swift}
In order to track the outburst in X-rays, we used a combination of instruments on board \textit{Swift}. To measure the day-to-day flux variability, we obtained data from the \textit{Burst Alert Telescope (BAT)}\footnote{Data is publicly available here: \url{https://swift.gsfc.nasa.gov/results/transients/weak/EXO1846-031.lc.txt}} which probes the `hard' X-ray energies, from 15--50\,keV. 

Observations were triggered as part of the ThunderKAT \textit{Swift/XRT} follow-up programme (PI: Motta), and obtained approximately every three days. The observations were made in window timing mode. Data were reduced using the \textit{Swift/XRT} product generator online reduction pipeline \citep{Evans2007,Evans2009}\footnote{\url{https://www.swift.ac.uk/user_objects/}} and downloaded for further analysis. The online reduction pipeline performs a standard analysis of XRT data and corrects for pile-up. The 0.6$-$10.0 keV energy band data were loaded into \textsc{XSPEC} 12.10.1 \citep*{XSpec} and fitted with three models: i) an absorbed power-law (\textsc{TBfeo} $\times$ \textsc{powerlaw}), ii) an absorbed disc model (\textsc{TBfeo} $\times$ \textsc{diskbb}) or iii) a combination of these two fits (\textsc{TBfeo} $\times$ (\textsc{powerlaw} + \textsc{diskbb})). We compare these three models and choose the best fit model with physical parameters based upon the $\chi$-squared values, but keeping the oxygen and iron abundances fixed to the defaults. Once the best fit model was obtained, absorbed and unabsorbed fluxes were extracted using the \textsc{cflux} task in the 1.0--10.0\,keV energy band, as well as the best spectral parameters and associated uncertainties. These spectral parameters and fluxes are reported in Table~\ref{tab:basicspeec}. 

\subsubsection{\textit{MAXI} data}\label{sec:MAXI}

\begin{figure*}
	\includegraphics[width=0.95\textwidth]{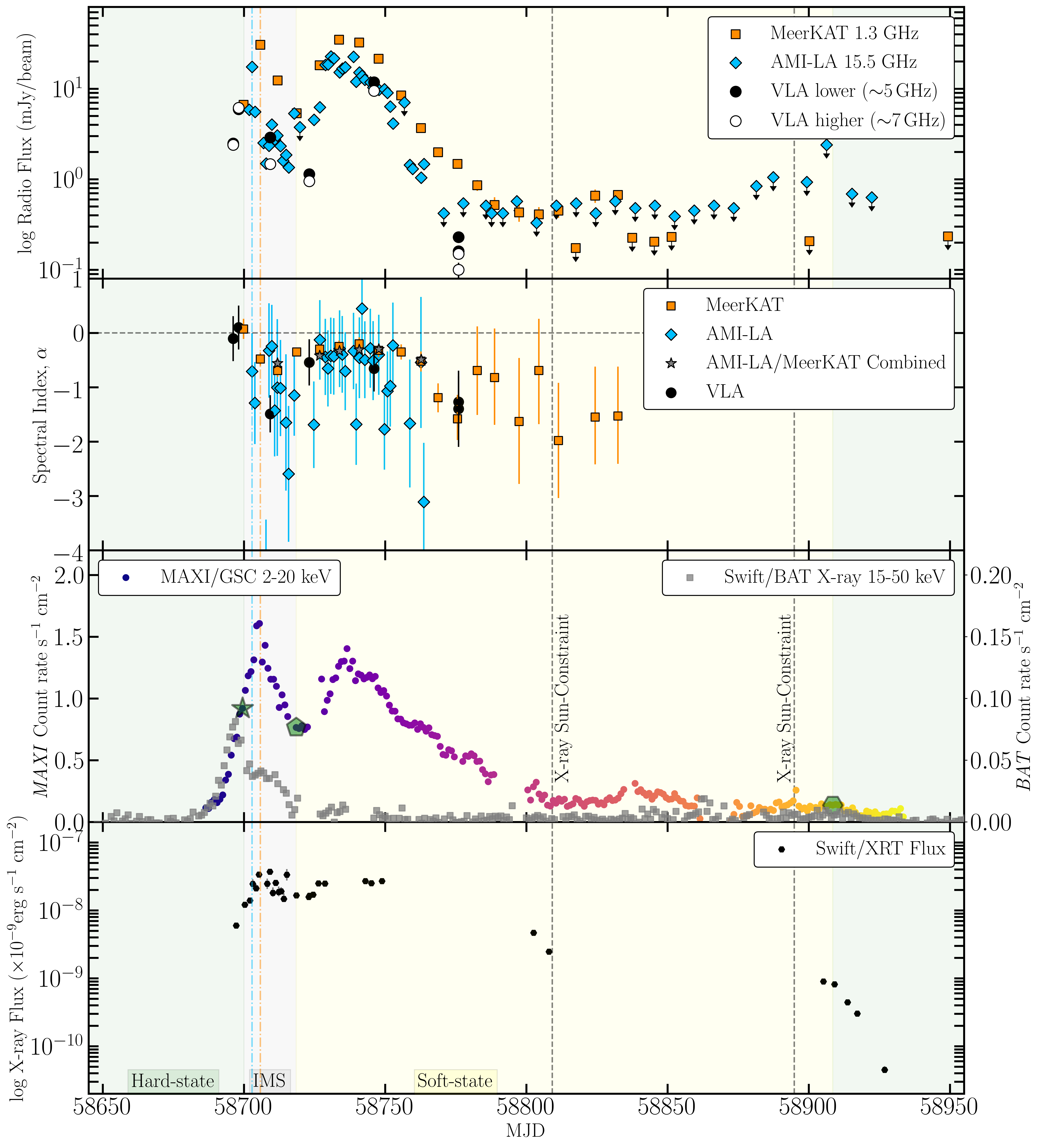}
    \caption{
    Radio and X-ray light curves and spectra of the 2019 outburst of the BHXB EXO 1846$-$031. The top panel shows the radio light curves for AMI-LA (blue diamonds), MeerKAT (orange squares) and VLA in the lower band near 5\,GHz (black circles) and upper band near 7\,GHz (black unfilled circles). Upper limits are denoted by downward facing arrows. The peaks in the AMI-LA and MeerKAT light curves are shown by blue or orange dashed lines in all panels, respectively. The second panel shows the in-band spectral indices for AMI-LA and MeerKAT in the same colour/marker style, with grey stars denoting the AMI-LA/MeerKAT combined spectral index and the two-band VLA spectral index denoted by black circles. In both of the top two plots, the final VLA observation shows two sources (see Table~\ref{tab:VLAdata}), and we include both of these sources on this plot.
    The bottom two panels show the X-ray data in order: the X-ray light curve of the outburst using all-sky monitor data from \textit{MAXI/GSC} and \textit{Swift/BAT}; the \textit{Swift/XRT} unabsorbed flux obtained from the X-ray spectral fits to the data. 
    For the all-sky monitor data, the grey squares show the X-ray count rate from \textit{Swift/BAT} in the 15--50\,keV `hard' band, while the coloured circles show the X-ray count rate in the \textit{MAXI/GSC} 2--20\,keV `soft' band. The colour for these circles denotes the day of observing, for direct comparison to Fig.~\ref{fig:HID}. 
    We represent the days where we consider the `hard' state transition as a green star in the fourth panel, the first point where the source is in the `soft' state as a green pentagon, and the green hexagon represents the first point where EXO 1846$-$031 is definitely back in the `hard' state: see discussion in Section 3.1. We label these the X-ray states in the bottom panel and shade the background green, grey or yellow for each of the X-ray states, respectively. We show with a black dashed line when EXO 1846$-$031 was Sun constrained for the \textit{Swift/XRT}.}
    \label{fig:lc}
\end{figure*}

\begin{figure*}
	\includegraphics[width=0.9\textwidth]{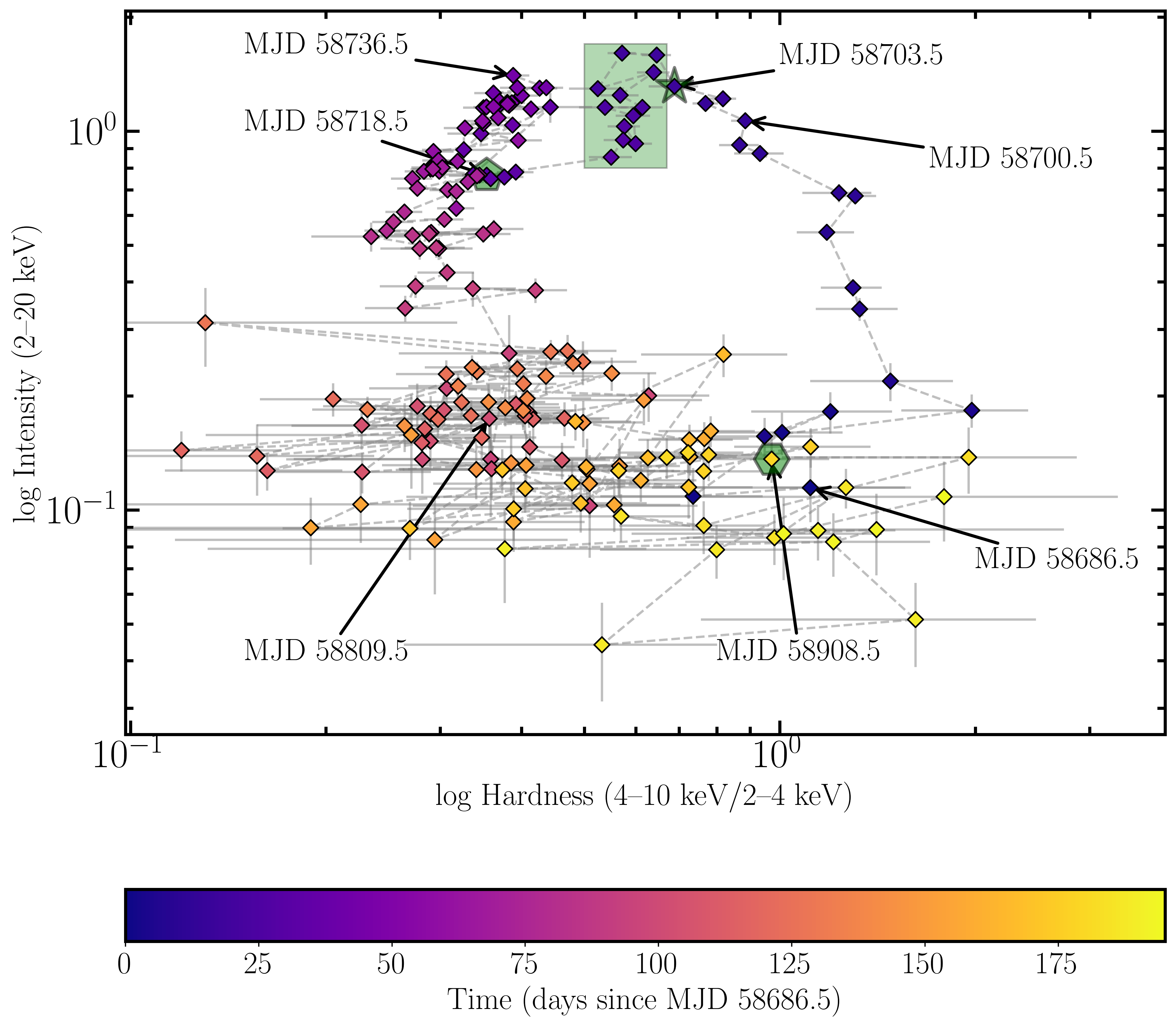}
    \caption{The \textit{MAXI/GSC} X-ray Hardness--Intensity diagram (HID) of the 2019 outburst of EXO1846, showing a characteristic q-shaped hysteresis of X-ray binaries in outburst. The data point colours correspond to the time of the observation, for direct comparison with Fig~\ref{fig:lc}, normalised to a day zero of MJD 58686.5. On this plot we label dates referring to important parts of the evolution of EXO 1846$-$031, moving in an anti-clockwise direction from the bottom right: MJD 58686.5 is defined as the zeroth day for the colour bar, MJD 58700.5 shows the point which we label as the exit of the `hard' state (see Section~\ref{sec:hardstate} and Fig.~\ref{fig:rmsint}, the green star indicates MJD 58703.5, %the last day of EXO 1846$-$031 in a `hard' state, 
    before transitioning into a `intermediate' state (green shaded box), MJD 58718.5 (green pentagon) denotes the first date when the source was unequivocally in the `soft' state, MJD 58736.5 is indicated as the approximate second X-ray light curve peak while in the `soft' state, MJD 58809.5 is shown as the point where EXO 1846$-$031 became sun constrained and the green hexagon (MJD 58908.5) indicates when EXO 1846$-$031 had returned to the X-ray `hard' state and faded into quiescence, although this exact date is unknown.
    }
    \label{fig:HID}
\end{figure*}

We obtained publicly available data from the Monitor of All-sky X-ray Image \citep[\textit{MAXI/GSC},][]{MAXI} for the duration of the outburst, which complemented the \textit{Swift} XRT data. The \textit{MAXI/GSC} data are available in different energy bands: 2--4, 4--10, 10--20 and 2--20\,keV. We plot the 2--20\,keV day-to-day counts in the fourth panel of Fig.~\ref{fig:lc} and plot the Hardness--Intensity Diagram (HID) of the 2019 outburst in Fig.~\ref{fig:HID}. We note that the \textit{MAXI} HID is also published in fig. 3 of \citet{Liu2020} with a detailed description of the changing X-ray classifications for the outburst using \textit{Insight/HXMT} data.

\begin{table*}
    \caption{\textit{Swift}-XRT X-ray spectra for each epoch, with fluxes and spectral parameters provided. The columns refer to: (1) The observation ID; (2) the MJD of the start of the observation; (3) the model used: \textsc{TBfeo}$\times$\textsc{powerlaw} denoted `$\alpha$', \textsc{TBfeo}$\times$\textsc{diskbb} denoted `$\beta$' and \textsc{TBfeo}$\times$(\textsc{diskbb} + \textsc{powerlaw}) denoted `$\delta$'.; (4) the absorbing column fit in the \textsc{TBfeo} model $\times$10$^{22}$ atoms cm$^{-2}$; (5) and (6) show the \textsc{diskbb} $T_{\rm in}$ at inner disk radius parameter in keV and the associated normalisation of this parameter; (7) and (8) show the fit powerlaw photon index ($\Gamma$) and associated normalisation; (9) the reduced $\chi$-squared of the fit and (10) the unabsorbed fluxes in the 1--10\,keV band $\times$10$^{-9}$, in erg cm$^{-2}$ s$^{-1}$, obtained using the \textsc{cflux} model in \textsc{XSpec}.The errors shown in this table are all at the 1$\sigma$ level.} 
    \renewcommand{\arraystretch}{1.2}
    \centering
    \begin{tabular}{c c c | c | c c | c c | c | c c}
    \hline  &  	 &   &   \multicolumn{1}{c}{\textsc{tbfeo}} &  \multicolumn{2}{c}{\textsc{diskbb}} &  \multicolumn{2}{c}{\textsc{powerlaw}}  & 
\multicolumn{1}{c}{red.} & Flux \\%& log Lum. \\ 
ObsID & MJD	& mod. &   $N_{\rm H}$ &   $T_{\rm in}$ &   norm. &  Phot.I. & norm. & $\chi^2$ & 1--10\,keV \\%& 1-10\,keV \\
    (1) & (2) & (3) & (4) & (5) & (6) & (7) & (8) & (9) & (10) \\%& (11) \\
    \hline
11500002 & 58697.259 & $\alpha$ & 3.85$_{-0.06}^{+0.06}$ & - & - & 1.52$_{-0.03}^{+0.03}$ & 0.88$_{-0.04}^{+0.04}$ & 1.0355 & 5.99$_{0.06}^{+0.06}$ \\%& 1.15$_{-0.01}^{+0.01}$ \\
11500003 & 58700.316 & $\alpha$ & 4.27$_{-0.04}^{+0.04}$ & - & - & 2.09$_{-0.02}^{+0.02}$ & 3.6$_{-0.1}^{+0.1}$ & 1.0197 & 12.1$_{0.1}^{+0.1}$ \\%& 2.38$_{-0.03}^{+0.01}$ \\
11500004 & 58702.094 & $\alpha$ & 4.47$_{-0.04}^{+0.04}$ & - & - & 2.13$_{-0.02}^{+0.02}$ & 4.4$_{-0.1}^{+0.1}$ & 1.074 & 14.0$_{0.1}^{+0.1}$ \\%& 2.58$_{-0.03}^{+0.03}$ \\
11500006 & 58703.099 & $\delta$ & 3.9$_{-0.3}^{+0.3}$ & 1.0$_{-0.1}^{+0.1}$ & 500$_{-300}^{+300}$ & 2.2$_{-0.6}^{+0.3}$ & 5$_{-4}^{+5}$ & 1.0491 & 24$_{-3}^{+4}$ \\%& 6.08$_{-0.71}^{+1.01}$ \\
11500007 & 58704.286 & $\delta$ & 3.8$_{-0.2}^{+0.3}$ & 0.99$_{-0.05}^{+0.06}$ & 800$_{-200}^{+200}$ & 2.0$_{-0.8}^{+0.4}$ & 2$_{-2}^{+4}$ & 1.0219 & 21$_{-2}^{+3}$ \\%& 5.00$_{-0.53}^{+0.93}$ \\
11500008 & 58705.288 & $\beta$ & 3.58$_{-0.05}^{+0.05}$ & 1.23$_{-0.02}^{+0.02}$ & 870$_{-50}^{+60}$ & - & - & 1.0875 & 33.6$_{-0.2}^{+0.5}$ \\%& 7.34$_{-0.10}^{+0.10}$ \\
11500009 & 58708.214 & $\delta$ & 4.2$_{-0.4}^{+0.3}$ & 1.1$_{-0.1}^{+0.2}$ & 300$_{-200}^{+300}$ & 2.8$_{-0.4}^{+0.3}$ & 10$_{-6}^{+7}$ & 1.0093 & 24$_{-4}^{+5}$ \\%& 6.59$_{-1.13}^{+1.26}$ \\
11500010 & 58709.201 & $\delta$ & 3.5$_{-0.2}^{+0.3}$ & 0.97$_{-0.05}^{+0.05}$ & 1100$_{-300}^{+200}$ & 1.8$_{-0.8}^{+0.6}$ & 2$_{-2}^{+5}$ & 1.1017 & 26$_{-2}^{+3}$ \\%& 4.92$_{-0.29}^{+0.60}$ \\
11500011 & 58710.264 & $\delta$ & 3.7$_{-0.3}^{+0.4}$ & 0.9$_{-0.1}^{+0.1}$ & 700$_{-400}^{+400}$ & 2.2$_{-0.7}^{+0.4}$ & 4$_{-3}^{+6}$ & 0.97435 & 18$_{-2}^{+4}$ \\%& 4.02$_{-0.43}^{+0.76}$ \\
11500012 & 58711.262 & $\delta$ & 4.1$_{-0.4}^{+0.7}$ & 0.9$_{-0.1}^{+0.3}$ & 1300$_{-1100}^{+800}$ & 2$_{-2}^{+1}$ & 5$_{-5}^{+19}$ & 0.92899 & 88$_{-1}^{+3}$ \\%& 5.09$_{-0.93}^{+1.87}$ \\
11500013 & 58712.325 & $\delta$ & 3.9$_{-0.4}^{+0.3}$ & 0.9$_{-0.1}^{+0.1}$ & 600$_{-300}^{+300}$ & 2.4$_{-0.6}^{+0.6}$ & 6$_{-4}^{+5}$ & 1.0135 & 19$_{-2}^{+4}$ \\%& 3.92$_{-0.51}^{+0.74}$ \\
11500014 & 58713.184 & $\delta$ & 3.7$_{-0.2}^{+0.3}$ & 1.04$_{-0.04}^{+0.05}$ & 800$_{-200}^{+100}$ & 1.9$_{-1.3}^{+0.7}$ & 1$_{-1}^{+4}$ & 1.0405 & 19$_{-1}^{+3}$ \\%& 3.98$_{-0.23}^{+0.56}$ \\
11500015 & 58714.112 & $\delta$ & 3.6$_{-0.2}^{+0.4}$ & 0.99$_{-0.05}^{+0.07}$ & 800$_{-200}^{+200}$ & 2$_{-2}^{+1}$ & 0.6$_{-0.6}^{+4.4}$ & 0.91189 & 15$_{-1}^{+3}$ \\%& 3.98$_{-0.23}^{+0.56}$ \\
11500016 & 58715.179 & $\delta$ & 4.6$_{-0.3}^{+0.3}$ & 1.0$_{-0.1}^{+0.2}$ & 400$_{-300}^{+400}$ & 3.2$_{-0.3}^{+0.3}$ & 21$_{-10}^{+12}$ & 1.206 & 33$_{-6}^{+7}$ \\%& - \\
11500017 & 58718.566 & $\beta$ & 3.71$_{-0.09}^{+0.09}$ & 0.89$_{-0.02}^{+0.02}$ & 1700$_{-200}^{+200}$ & - & - & 0.87492 & 16.60$_{0.05}^{+0.02}$ \\%& - \\
%11500018 & - & - & - & - & - & - & - & - & - & - \\
88981001 & 58722.957 & $\beta$ & 3.8$_{-0.1}^{+0.1}$ & 0.86$_{-0.02}^{+0.02}$ & 1900$_{-300}^{+300}$ & - & - & 1.0751 & 15.77$_{-0.07}^{+0.08}$ \\%& - \\
88981002 & 58723.157 & $\beta$ & 3.59$_{-0.08}^{+0.08}$ & 0.93$_{-0.02}^{+0.02}$ & 1500$_{-100}^{+100}$ & - & - & 1.1268 & 16.38$_{-0.04}^{+0.05}$ \\%& - \\
11500019 & 58724.481 & $\beta$ & 3.82$_{-0.07}^{+0.07}$ & 0.89$_{-0.01}^{+0.01}$ & 1800$_{-100}^{+100}$ & - & - & 1.0124 & 16.97$_{-0.04}^{+0.04}$ \\%& - \\
11500020 & 58726.342 & $\beta$ & 3.68$_{-0.05}^{+0.06}$ & 0.91$_{-0.01}^{+0.01}$ & 2400$_{-100}^{+160}$ & - & - & 1.0984 & 25.02$_{-0.04}^{+0.05}$ \\%& - \\
11500021 & 58728.665 & $\beta$ & 3.72$_{-0.05}^{+0.05}$ & 0.92$_{-0.01}^{+0.01}$ & 2200$_{-100}^{+100}$ & - & - & 1.1713 & 24.76$_{-0.04}^{+0.04}$ \\%& - \\
%\hline
%11500022 & - & - & - & - & - & - & - & - & - & - \\
11500024 & 58743.002 & $\beta$ & 3.79$_{-0.07}^{+0.08}$ & 0.93$_{-0.01}^{+0.01}$ & 2400$_{-200}^{+200}$ & - & - & 1.016 & 26.9$_{-0.6}^{+0.7}$ \\%\\%& - \\
11500025 & 58745.139 & $\beta$ & 3.74$_{-0.06}^{+0.06}$ & 0.97$_{-0.01}^{+0.01}$ & 1800$_{-100}^{+100}$ & - & - & 1.0299 & 25.1$_{-0.5}^{+0.5}$ \\%\\%& - \\
11500026 & 58748.839 & $\beta$ & 3.99$_{-0.09}^{+0.09}$ & 0.89$_{-0.01}^{+0.01}$ & 2900$_{-300}^{+300}$ & - & - & 1.1024 & 26.9$_{-0.8}^{+0.8}$ \\%\\%& - \\
11500028 & 58802.562 & $\beta$ & 2.8$_{-0.1}^{+0.2}$ & 0.88$_{-0.03}^{+0.03}$ & 500$_{-100}^{+100}$ & - & - & 0.98232 & 4.7$_{-0.3}^{+0.3}$ \\%\\%& - \\
11500029 & 58808.015 & $\beta$ & 3.6$_{-0.1}^{+0.1}$ & 0.76$_{-0.02}^{+0.02}$ & 550$_{-70}^{+80}$ & - & - & 1.2375 & 2.5$_{-0.1}^{+0.1}$ \\%\\%& - \\
11500030 & 58905.157 & $\alpha$ & 3.4$_{-0.1}^{+0.1}$ & - & - & 2.15$_{-0.07}^{+0.07}$ & 0.29$_{-0.03}^{+0.03}$ & 0.96611 & 0.90$_{0.03}^{+0.04}$ \\%\\%& - \\
11500031 & 58909.125 & $\alpha$ & 3.5$_{-0.2}^{+0.2}$ & - & - & 2.20$_{-0.09}^{+0.09}$ & 0.28$_{-0.03}^{+0.04}$ & 1.0726 & 0.82$_{0.04}^{+0.04}$ \\%\\%& - \\
11500032 & 58913.774 & $\alpha$ & 2.8$_{-0.2}^{+0.2}$ & - & - & 1.9$_{-0.1}^{+0.1}$ & 0.11$_{-0.02}^{+0.02}$ & 0.99995 & 0.44$_{0.02}^{+0.02}$ \\%\\%& - \\
11500033 & 58917.165 & $\alpha$ & 3.0$_{-0.2}^{+0.2}$ & - & - & 1.7$_{-0.1}^{+0.1}$ & 0.06$_{-0.01}^{+0.01}$ & 0.9711 & 0.30$_{0.01}^{+0.03}$ \\%\\%& - \\
\hline
    \end{tabular}
    \label{tab:basicspeec}
\end{table*}

\section{Results}

The radio and X-ray light curves of the 2019 outburst of the BHXB candidate EXO 1846$-$031 are presented in Fig.~\ref{fig:lc}, with a zoom in of the radio peaks in Fig.~\ref{fig:lc_appendix}. The top panel shows the radio light curves for all three arrays and the second panel shows the radio spectral indices obtained from in-band and inter-array quasi-simultaneous observations (see Section~\ref{sec:specindex}). The third panel in Fig.~\ref{fig:lc} shows the X-ray monitor (\textit{Swift/BAT} and \textit{MAXI/GSC}) count rates and the lower panel shows the X-ray fluxes obtained from the \textit{Swift/XRT} X-ray spectral fitting described in Section~\ref{sec:xrayobs}. We also compute a Hardness--Intensity Diagram (HID, Fig.~\ref{fig:HID}), a rms--Intensity diagram (Fig.~\ref{fig:rmsint}) and first analyse the X-ray spectral states (Section~\ref{sec:xraydisc}). Finally, we describe the radio behaviour and source evolution (Section~\ref{sec:radiodisc}), including presenting the VLA images (Fig.~\ref{fig:radioejecta}).

\label{sec:lightcurves}

\subsection{The X-ray spectral states and Hardness--Intensity diagram}
\label{sec:xraydisc}

Broadly, the outburst tracked the standard `q-shaped' pattern of X-ray binaries in outburst: initially the source resided in the spectral `hard' state (Section~\ref{sec:hardstate}) from MJD 58687 at low fluxes, before reaching a flux peak around MJD 58700 and then softening into the `intermediate' states (IMS) (Section~\ref{sec:hardintstate}). EXO 1846$-$031 remained in the IMS (MJD 58700--58717) before entering a `soft' state for a prolonged period and gradually faded (after MJD 58717, see Section~\ref{sec:softstate}). Finally, at some point after MJD 58887, the source returned to the `hard' state. We describe each of these states in detail below, including their specific X-ray behaviour.

\subsubsection{The X-ray Spectral `Hard' State: MJD 58687--58700}
\label{sec:hardstate}

At the start of the outburst, EXO 1846$-$031 was in the X-ray spectral `hard' state, as evidenced by the power-law dominated X-ray spectra and the steep-rise on the right hand side of the HID (Fig.~\ref{fig:HID}). The exact end of the `hard' state is not clear, though \textit{Insight/HXMT} spectra indicate a transition between MJD 58697 and MJD 58703 \citep{Liu2020}.

\begin{figure}
    \centering
    \includegraphics[width=\columnwidth]{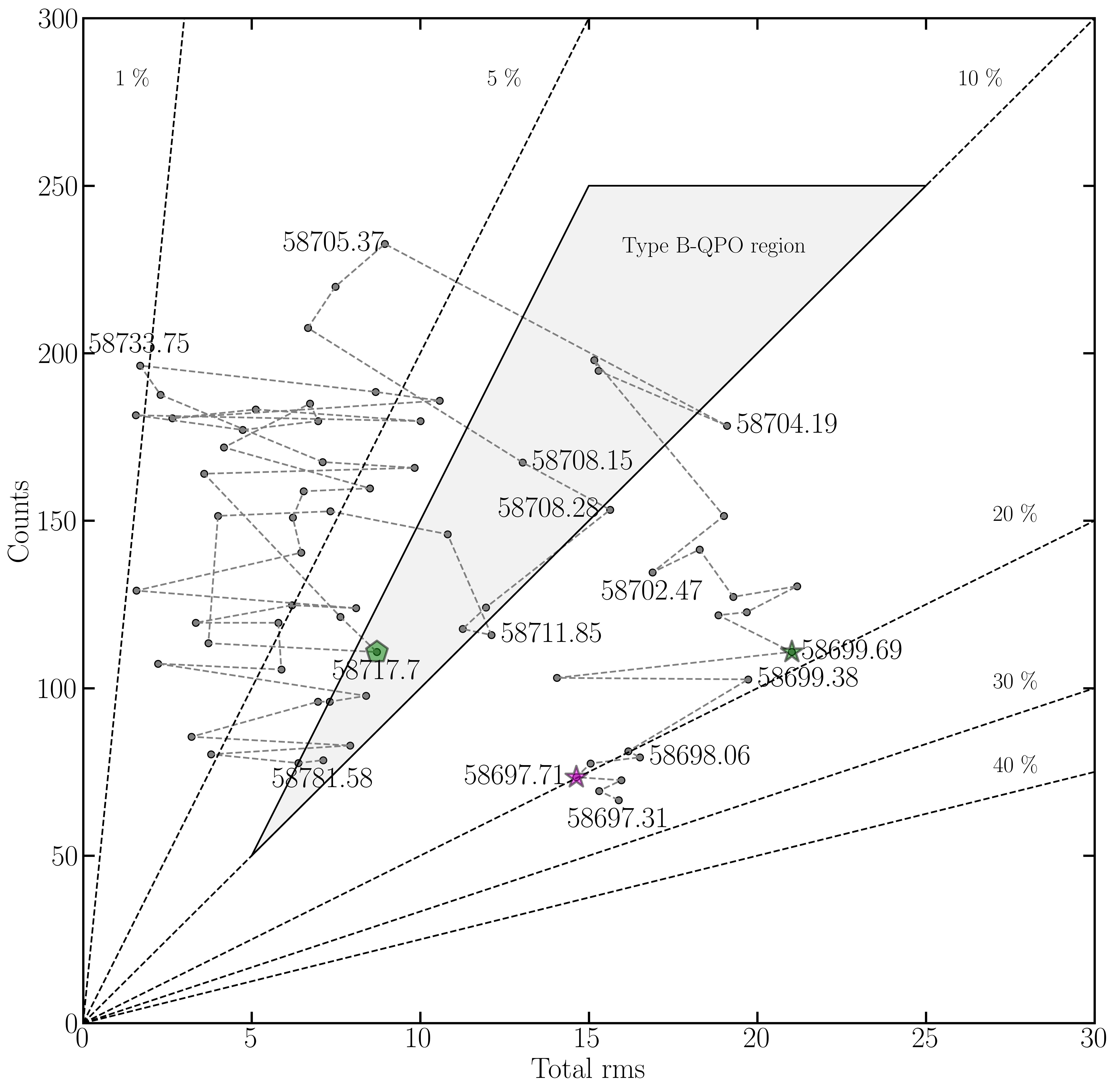}
    \caption{RMS intensity diagram for the 2019 Outburst of EXO 1846$-$031, using digitised \textit{Insight/HXMT} data from \citet[][]{Liu2020}. MJDs are labelled for interesting points of the outburst. The magenta star indicates where \citet[][]{Liu2020} suggest the source transitioned from the `hard' to the `hard-intermediate' state on MJD 58697.
    The green star (as in Fig.~\ref{fig:lc}) is the last data point which is moving along the 20 per cent fractional rms variability line, and is dated on MJD 58699.7. After this date the source clearly departs and never returns, so we consider that $\sim$MJD 58700 is likely the point where the state transition occurred. There is one anomalous data point on MJD 58699.5, i.e. the data point preceding the green star, but this would require a significant change in the state in less than 4 hours, between the two data points and does not signify a state transition. The source proceeds to the 10 per cent variability line and departs across it around MJD 58704.19. The region shaded in light grey is the region identified by \citep[][]{munozdarias-rmsint} as the region where Type-B QPOs were seen in GX339$-$4. Whilst a couple of data points exist in this region, they are very short lived, often being only an hour or two long. It is therefore possible that the times for the Type-B QPO to be sampled were not available to the data in the outburst at significant signal to noise ratio, and as such explains why Type-B QPOs were not observed. This figure is replicated in Figure~\ref{fig:lc_appendix} but zoomed in on the two peaks in the radio and X-ray light curves. 
    }
    \label{fig:rmsint}
\end{figure}

We digitised the \textit{Insight/HXMT} data shown in fig. 1 of \citet[][]{Liu2020}, and plotted the rms-intensity diagram \citep[][]{munozdarias-rmsint} in Fig.~\ref{fig:rmsint} to ascertain a better transition date: the rms-intensity diagram is a good tracer of the exit from the `hard' X-ray spectral state and shows that the divergence from the `hard line' may have happened around MJD 58700 instead. For the purposes of this work, we consider the `low-hard' state to finish on MJD 58700, i.e. when the source leaves the `hard line' of the rms-intensity diagram. We label this point with a green star in panel 3 of Fig.~\ref{fig:lc} and Fig.~\ref{fig:HID}. 

Having defined the X-ray spectral `hard' state, i.e. up to MJD 58700, we can see that the X-ray flux in the \textit{Swift/BAT} 15--50\,keV data was decreasing after MJD 58700, but the \textit{MAXI/GSC} 2--20\,keV flux was still increasing. The two X-ray spectral fits from the \textit{Swift/XRT} data around this time also show an increase in flux from (6--12)$\times$10$^{-9}$ erg cm$^{-2}$ s$^{-1}$ between the MJD 58697 and MJD 58700, and are best described as an absorbed power-law, although the photon index steepens between the two observations from 1.52 to 2.09.

\subsubsection{The X-ray Spectral `Intermediate' States: MJD 58700--58717}
\label{sec:hardintstate}

The sharp turn in the rms-intensity diagram after MJD 58700 indicates that EXO 1846$-$031 transitioned into the `intermediate state' (IMS), on this date. We show the IMS duration in Fig.~\ref{fig:HID} as a shaded green region. During this time-frame the \textit{Swift/XRT} X-ray spectra change from an absorbed power-law model to an absorbed multicolour disk-blackbody plus powerlaw model, with a gradually increasing contribution to the flux from the disk component. The peak in \textit{MAXI/GSC} is on MJD 58703, though there is some day-to-day variability in the X-ray flux extracted from the \textit{Swift/XRT} spectra. After MJD 58703, the X-ray fluxes appear to decrease in all X-ray instruments until approximately MJD 58720. 

Between MJD 58711 and MJD 58717, \citet{Liu2020} suggest that the source is in the `soft intermediate state' (SIMS), though no Type-B QPO is observed - a common feature of the SIMS. We do not make the distinction between the `hard' and `soft' intermediate states, though the rms-intensity plot (Fig.~\ref{fig:rmsint}) shows that the \textit{Insight/HXMT} observations traverse through the region where Type-B QPOs appear briefly on several occasions. Hence it is possible that the SIMS was never fully sampled. Therefore, we consider that EXO 1846$-$031 was in some form of intermediate state from  MJD 58700 to MJD 58717, but it is not possible to say whether the SIMS was sampled.

\subsubsection{The X-ray Spectral `Soft' State: after MJD 58717}
\label{sec:softstate}
After MJD 58718, the \textit{MAXI/GSC} light curve begins to increase once more, reaching a peak around MJD 58736. However, the increase in count rate is not mirrored in the 15--50\,keV \textit{Swift/BAT} light curve. The X-ray spectra after MJD 58718 no longer require the presence of a power-law to provide adequate fits to the data, which is best described by an absorbed multicolour disk model. From this point onwards, EXO 1846$-$031 is unequivocally in the X-ray `soft' state, shown by the green pentagon in panel 3 of Fig.~\ref{fig:lc} and Fig.~\ref{fig:HID}. From MJD 58736 onwards, EXO 1846$-$031 starts to decrease in X-ray count rate until $\sim$MJD 58810. There are not many observations with \textit{Swift/XRT} during this period, although the flux clearly decrease from $\sim$25$\times$10$^{-9}$ erg cm$^{-2}$ s$^{-1}$ around MJD 58745 to $\sim$3$\times$10$^{-9}$ erg cm$^{-2}$ s$^{-1}$ around MJD 58805. 

EXO 1846$-$031 became Sun-constrained for \textit{Swift/XRT} between MJD 58809 until MJD 58894. Therefore, it is unclear when EXO 1846$-$031 returned to the X-ray `hard' state. 
The \textit{Swift/XRT} data from MJD 58905 onwards show a source dominated by a fading power-law, so the transition back into the `hard' state must have started before MJD 58906. We place a green hexagon at this date in the X-ray light curve (Fig.~\ref{fig:lc}) and HID (Fig.~\ref{fig:HID}).

\subsection{Radio light curves, spectra and evidence for jets}
\label{sec:radiodisc}
We now consider the radio data and compare it to the X-ray data and the state transition dates. 

\begin{figure}
	\includegraphics[width=0.95\columnwidth]{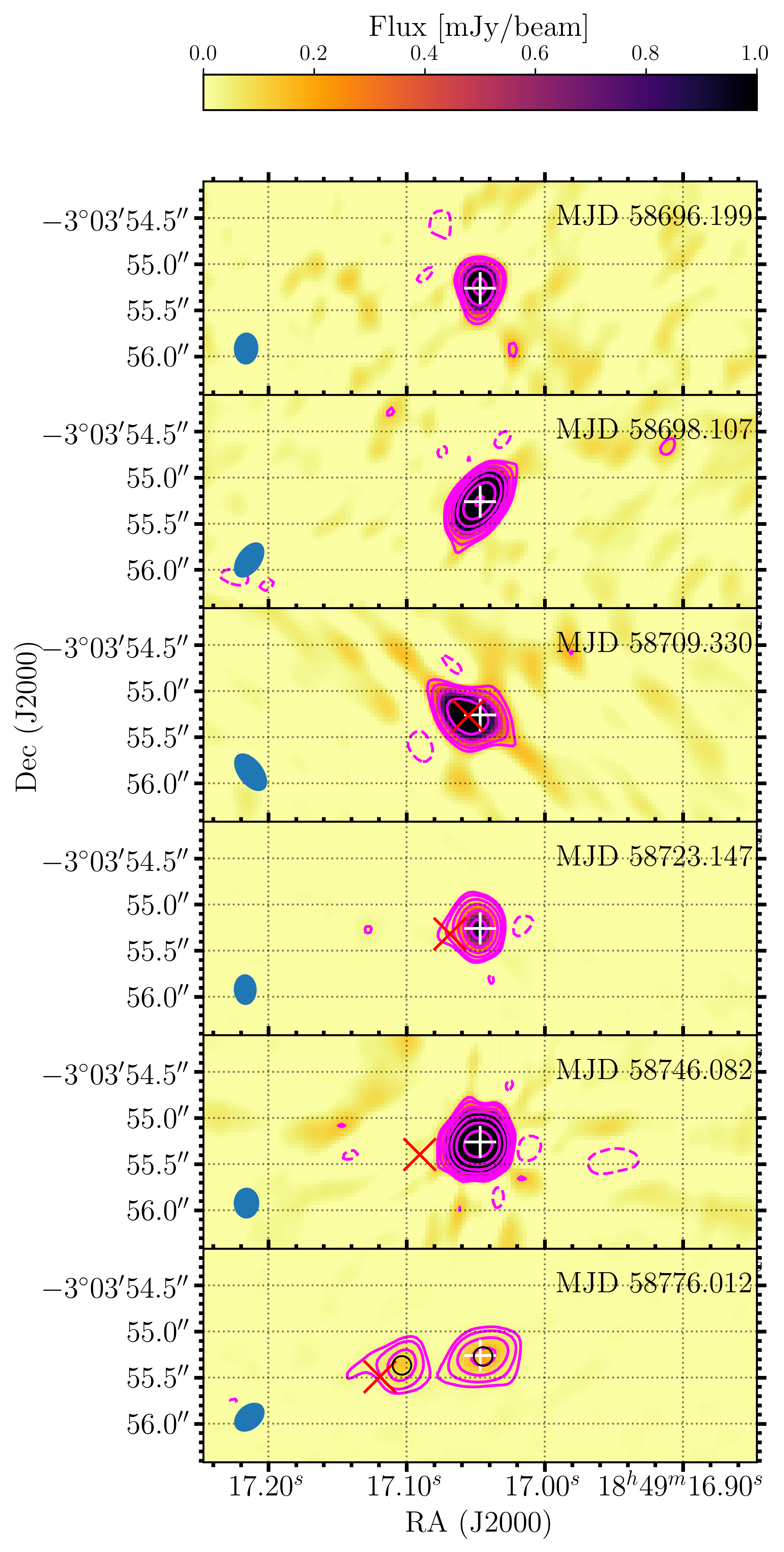}
    \caption{All six VLA epochs of the 2019 outburst of EXO 1846$-$031 in chronological order from top to bottom. The date of the observation is labelled in the top right hand corner of each panel: the first two observations were while the source was in the X-ray `hard' state, the third was during the `intermediate' state, and the final three observations while the source was in the `soft' spectral state. Each panel shows the full-band VLA dataset, but fluxes for each sub-band are given in Table~\ref{tab:VLAdata}. The background colour scale is the same for all epochs and labelled at the top of the figure, but the contour levels are set as the image rms $\times-3, 3, 5, 10, 20, 40, 80, 160, 320$. The image rms noise levels for the full-band width used above are shown in Table~\ref{tab:VLAdata}. The synthesized beam size and shape is shown in the bottom left-hand corner. The white `plus' sign indicates the position of EXO 1846$-$031 as ascertained from the first dataset and \citet{Miller-JonesATel}. The red cross in panel 3 denotes the position of the first jet component, and in all later panels this red cross denotes the position where this source should reside, given a jet ejection date around MJD 58700, assuming a constant velocity. The two black circles in the final panel show the positions of the two components detected in this final epoch, showing that the eastern most component has not moved as far eastwards as expected, and the central component is located slightly westwards of the core position and therefore may be receding jet.
    }
    \label{fig:radioejecta}
\end{figure}

The first two VLA epochs were observed while the X-ray emission was still rising during the X-ray `hard' state (Section~\ref{sec:hardstate}), and show a compact radio source with an increasing radio flux from $\sim$2.5\,mJy on MJD 58696 to $\sim$6\,mJy on MJD 58698 (see top two panels in Fig.~\ref{fig:radioejecta}). While EXO 1846$-$031 was in the X-ray spectral `hard' state, only one other radio observation with MeerKAT was obtained on MJD 58699.839. During this time, both the in-band VLA and MeerKAT observations show a flat ($\alpha \sim 0$) radio spectral index, as expected for sources in the `hard' state as the radio-flux increases but the radio emission remains optically-thick.

During the IMS (Section~\ref{sec:hardintstate}, MJD 58700--58717), we observe significant changes in the AMI-LA and MeerKAT light curves and spectra, but we only have one VLA data point (third panel of Fig.~\ref{fig:radioejecta}) during this phase.
The MeerKAT and AMI-LA light curves both show peaks in their light curves at a similar time to the \textit{MAXI} data (around MJD 58703), denoted in Fig.~\ref{fig:lc} with a vertical blue or orange dot-dashed line, respectively. A zoom in of this region is shown in Figure~\ref{fig:lc_appendix}. The AMI-LA data show a large day-to-day variability around this peak: on MJD 58702 the peak flux density is 17.4$\pm$1.4\,mJy beam$^{-1}$, but on MJD 59701 and 58703, it is $\sim$5$-$6\,mJy beam$^{-1}$. We label this fast rise as the `first radio flare', which we will discuss further in Section~\ref{sec:flares}. We split the AMI data on MJD 58702 into three parts and find no obvious intra-observation variability in this epoch. 
After the first peaks, all radio light curves decrease in flux density, with some AMI-LA epochs around MJD 58710 decreasing to $\sim$1\,mJy beam$^{-1}$. Around this time the MeerKAT flux decreases to $\sim$5\,mJy beam$^{-1}$. A single VLA epoch (MJD 58709.330) during this time period shows a spectrally-steep source ($\alpha=-1.5\pm0.3$) that has a centroid that is spatially-offset by 0.13-arcsec eastwards from the radio source identified in the first VLA epoch (see third panel in Fig.~\ref{fig:radioejecta}). 
We note that the error in the source position is 10 per cent the VLA synthesized beam, i.e. $\sim$0.05-arcsec. Therefore, we made a uniform weighted image of this epoch and find a source at the same position, but with a smaller synthesized beam ($0.35\times0.27$-arcsec), which provides a smaller positional uncertainty ($\sim$0.04 arcsec) and also confirms this positional movement. But, folding the positional uncertainties in quadrature from the initial BH position from \citet{Miller-Jones2009}, we obtain an error of the third epoch centroid of 0.05-arcsec, which represents a $\sim$2.5$\sigma$ movement eastwards. We attempted to force-fit a second component at the nominal position of the BH, but the fit was unconstrained. Thus, we consider this detection to be the first sign of a jet ejection in this system, despite the low significance. 
The radio spectral indices are consistent with $\alpha \la$ 0 throughout this period, but tend to more negative values by MJD 58717, at the end of the IMS. 

As EXO 1846$-$031 entered the X-ray `soft' state ($\sim$MJD 58718, see Section~\ref{sec:softstate}), the radio light curve mirrors the increase in the \textit{MAXI/GSC} X-ray count rate, with secondary AMI-LA/MeerKAT light curve peaks  observed between MJD 58731 to MJD 58739. In particular the AMI-LA light curve increases from $4.7\pm0.4$ on MJD 58724.811 to $22.9\pm1.8$\,mJy beam$^{-1}$ on 58730.811, which we label the `second radio flare' (see Section~\ref{sec:flares}). These second radio peaks are broader than the initial peak in the radio light curves. A single VLA observation on MJD 58723 during this flux density increase (panel four in Fig.~\ref{fig:radioejecta}) shows a faint ($\sim$1\,mJy beam$^{-1}$) source that is slightly extended eastwards, has a spectral index of $-$0.5$\pm$0.4 and is spatially consistent with the initial detections of the core (see Section~\ref{sec:JetSpeed} for a discussion). After MJD 58740, the radio flux densities decreased across all frequencies over the following two months. The last detections were recorded on MJD 58763.738 (AMI-LA), and MJD 58832.427 (MeerKAT). Two VLA epochs during this phase (MJD 58746 and 58776, bottom two panels in Fig.~\ref{fig:radioejecta}) show resolved source structures, with the sixth epoch showing a clear component 0.85-arcsec eastwards of the nominal BH position. In addition, the sixth VLA epoch showed a source offset by 0.04-arcsec westwards from the nominal BH position but we stress that this offset is within the positional uncertainty of the initial radio detection (see discussion in Section~\ref{sec:JetSpeed} about the nature of this radio emission). The radio emission during this period is optically-thin, i.e. a radio spectral index $< 0$, in all in-band observations across all observatories. After MJD 58750 the radio spectra become increasingly steeper, until the source is no longer detected. 

Finally, we note that around the time EXO 1846$-$031 returned to the X-ray `hard' state (before MJD 58906), no quasi-simultaneous sub-mJy noise radio detections were made (see Section~\ref{sec:radiodisc}); the last sub-mJy noise AMI-LA  observation occurred on MJD 58887. 
Therefore, we suggest the return to the `hard' state may have happened between MJD 58887 and MJD 58905.

\section{Discussion}

We now discuss our results in the context of X-ray binaries, focusing on source parameters like the distance (Section~\ref{sec:distance}), the radio:X-ray plane (Section~\ref{sec:RXplane}) and the radio jets detected in the VLA data (Section~\ref{sec:JetSpeed}).

\subsection{On the distance to EXO 1846--031}
\label{sec:distance}
Given the lack of optical counterpart \citep{Parmar1993}, the distance to EXO 1846$-$031 is not well constrained. The current best estimate is $\sim$7\,kpc, but this distance is calculated assuming the peak flux in \textit{EXOSAT} measurements corresponds to a luminosity of 10$^{38}$ erg\citep{Parmar1993}. An accurate distance measurement is necessary for calculating the source luminosity for scaling relationships such as the radio:X-ray plane \citep[][see Section~\ref{sec:RXplane}]{GalloFenderPooley,Corbel2003,Coriat2011}.

A renewed search for an optical counterpart was made for the 2019 outburst, but no optical counterpart was found in the Zwicky Transient Facility (ZTF) to a limiting magnitude of $r \sim 23$ mag band \citep{BellmATel}. A search was also performed using the LCOGT network (Dave Russell, priv comm) with a limit in the $i'$ band magnitude of 21.7. The large extinction in the direction of EXO 1846$-$031 ($N_{\rm H} \approx 3.5 \times 10^{22}$
cm$^{-22}$, \citealt{Parmar1993}), which is confirmed by our X-ray spectral fits, likely accounts for the non-detections in both the archival and new data.

We can attempt to place limits on the distance by searching for absorption in the 21\,cm \ion{H}{i} line against background sources \citep[see e.g.][for a recent examples of \ion{H}{i} distance measurements to the XBs MAXI J1535$-$571 and MAXI J1348$-$630]{Chauhan2019,Chauhan2021}. We combined the MeerKAT data for the three brightest epochs ($\ga$ 30\,mJy beam$^{-1}$) and extracted the radio spectrum around the source position. We compared this spectrum to a nearby distant source to assess the contribution of \ion{H}{i} in the line-of-sight to EXO 1846$-$031, but, there was no detection of the \ion{H}{i} line above a 3$\sigma$ threshold and so no estimate of the distance can be made from these data. 

The state transition luminosity distribution of BHXBs \citep[][]{Maccarone2003,Kalemci2013,VahdatMotlagh2019} can be used as a standard method to constrain the distance in many BHXBs \citep[]{Homan2005,Miller-Jones2012,Saikia2022}. During the transition from the soft to the hard state, BHXBs generally have a mean Eddington luminosity fraction of 1.9$\pm$0.2 per cent \citep[][]{Maccarone2003}, and a more conservative range of 0.3--3 per cent of Eddington luminosity \citep[][]{Kalemci2013}, but see also the example cases of 4U 1630$-$47 and MAXI J1535$-$571, which underwent transitions at 0.008 per cent \citep[][]{Tomsick2014} and $\sim$0.002 per cent \citep[e.g.][]{Chauhan2019} the Eddington luminosity, respectively. For EXO 1846$-$031, although the exact date of the soft to the hard state transition is not certain, it appears to have happened around MJD 58905 when the source was detected post Sun-Constraint. During this epoch, the 1--10 keV X-ray flux with \textit{Swift}/XRT is 0.9$\times$10$^{-9}$ erg cm$^{-2}$ s$^{-1}$ (see Table~\ref{tab:basicspeec}). Assuming a black hole mass of 3.24$\pm$0.2 M$_{\odot}$ \citep[][]{Strohmayer,Wang2021}, 1.9$\pm$0.2 per cent of Eddington luminosity \citep[][]{Maccarone2003} and a bolometric correction factor of 2 relative to the \textit{Swift}/XRT band, we obtain a distance of $\sim$6.0$\pm$0.3\,kpc for the source. For the more conservative range of 0.3--3 per cent Eddington luminosity \citep[][]{Kalemci2013}, we constrain a distance range of 2.4--7.5\,kpc.

\begin{figure*}
	\includegraphics[width=0.9\textwidth]{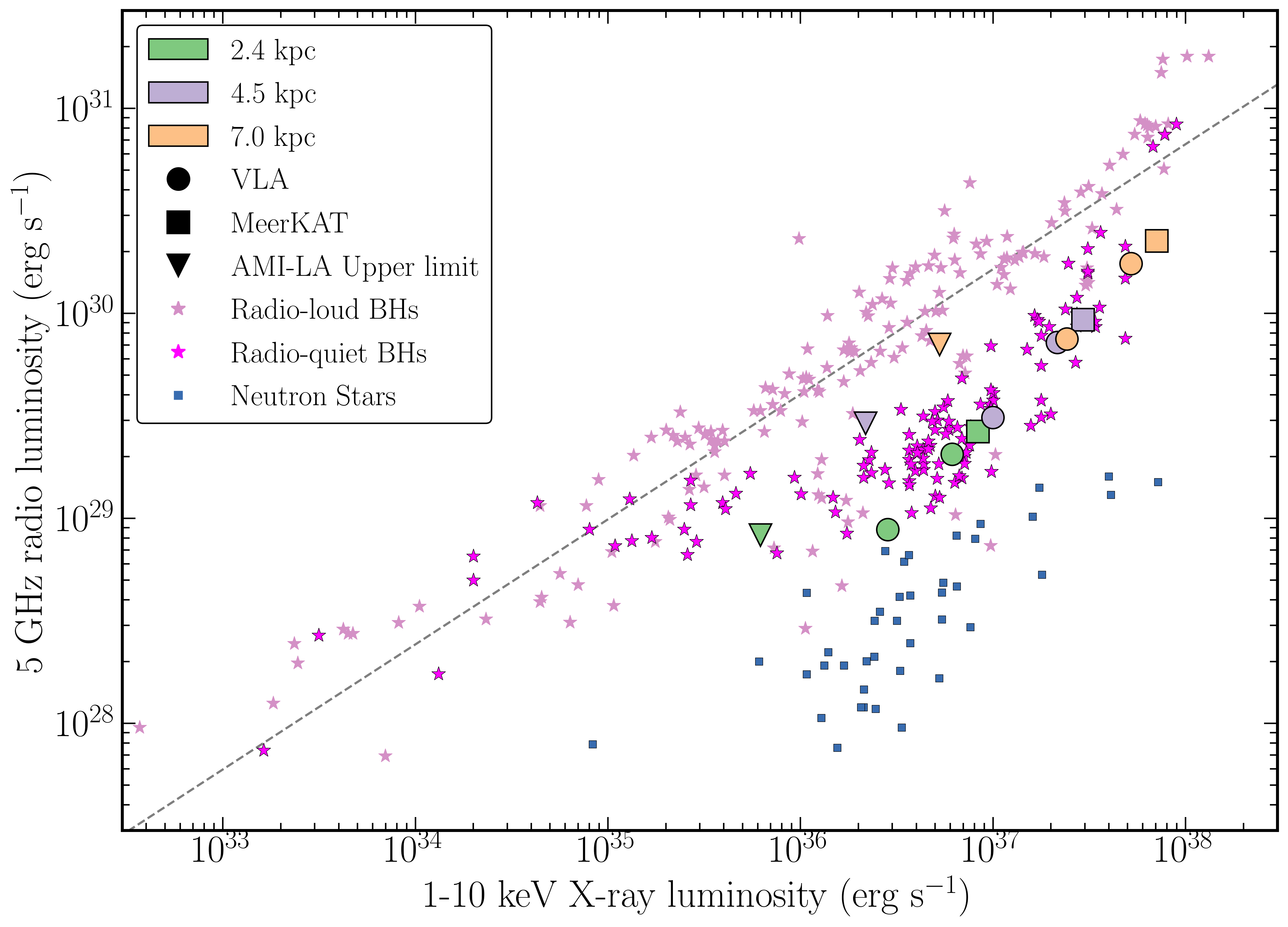}
    \caption{The radio:X-ray plane of the BHXB EXO 1846$-$031 for the 2019 outburst, using the VLA (circles) and MeerKAT (squares) observations with the 1--10\,keV flux densities from \textit{NICER} and \textit{MAXI/GSC} satellites (see text). Included is an additional data point from an upper limit in the AMI-LA observations after the source had transitioned back into the `hard' state on MJD 58905, shown as a downward triangle.
    Only observations with quasi-simultaneous radio and X-ray data are included in this plot. The EXO 1846$-$031 data points are scaled to three different distances: 2.4, 4.5 and 7.0 kpc (based off the discussion in  Section~\ref{sec:distance}). The pale and bright pink stars in the background show a compilation of `radio-loud' and `radio-quiet' objects from \citet[][]{arash_bahramian_2018_1252036}, respectively. The purple squares are neutron stars from the same compilation, showing that they are unlikely to be a good fit to the EXO 1846$-$031 data presented here.
    }
    \label{fig:LrLx}
\end{figure*}

\subsection{The radio:X-ray Plane}
\label{sec:RXplane}

Given the lack of precise distance measurement, we attempt to reduce this uncertainty by plotting the radio and X-ray data points for the 2019 outburst on the radio:X-ray plane in Fig.~\ref{fig:LrLx}, using three different distances: 2.4\,kpc (the bottom of our distance range from Section~\ref{sec:distance}), 4.5\,kpc and the literature value, 7.0\,kpc. For a given source distance, an object will move diagonally up and down the radio:X-ray plane, but as the relation between the two parameters is non-linear, an object will appear to fit the underlying distributions differently as a function of distance. We compiled all of the `hard' state (before MJD 58700)  quasi-simultaneous (i.e. within $\sim$1 day) data presented here for which the correlation holds. This limits our data set to two VLA data points (MJD 58696.199 and 58698.107) and one MeerKAT data point (MJD 58699.839).

We compare the first VLA data point to \textit{NICER} X-ray observations from 15~h prior \citep{BultATel}, as originally performed by \citet[][]{Miller-JonesATel}.
For the second VLA data point and MeerKAT data point, we extracted the X-ray spectrum from the \textit{MAXI/GSC} instrument in the 1--10\,keV energy range using the online on-demand extraction tool\footnote{\url{http://maxi.riken.jp/mxondem/}}, fitting the spectrum using \textsc{XSpec} with an absorbed power-law model and an average absorbing column from our \textit{Swift/XRT} fits of 4$\times$10$^{22}$ atoms cm$^{-2}$. For the second VLA data point, we calculated a flux of 0.9$\times$10$^{-8}$ erg cm$^{-2}$ s$^{-1}$. We calculate a flux of 1.2$\times$10$^{-8}$ erg cm$^{-2}$ s$^{-1}$ at the time of the MeerKAT epoch, consistent with the \textit{Swift/XRT} fluxes. We scale the MeerKAT radio flux density to 5\,GHz using the in-band spectral index ($\alpha = 0.1$). 

Finally, we also plot the \textit{Swift/XRT} data from MJD 58905 with the upper limit from our AMI-LA data on MJD 58906, as we are confident that the source had transitioned back to the `hard' state at the end of the outburst by this date. 

Regardless of the distance, the source appears under-luminous in the radio band to be considered one of the `radio-loud' BHXBs, but a simple linear fit for the three `hard' state data points gives a gradient of $\sim$1.0, which is in between the values for the `radio-quiet' and `radio-loud' populations \citep[and the neutron stars see e.g.][]{Coriat2011}. 

As for the distance, the 4.5\,kpc value aligns well with the under-lying `radio-quiet' BH population, whereas the brightest data points for the 7\,kpc literature value are X-ray brighter than many other previous XB or NS detected for any given radio luminosity. Given the paucity of data points, EXO 1846$-$031 could also be adequately fit at a distance of 2.4\,kpc, so it is not possible to constrain the distance further than the estimate for the soft-to-hard transition noted in Section~\ref{sec:distance}. 
The NSs are also plotted on this diagram, but EXO 1846$-$031 would need to be among one of the radio-brightest NSs detected, at any distance, given the data, including our lowest distance of 2.4\,kpc. This point was also discussed by \citep{Miller-JonesATel}, who stated that the source would have to be closer than 3.7\,kpc to be a NS. 

Therefore, we suggest that EXO 1846$-$031 is a `radio-quiet' BH that lies somewhere between 2.4\,kpc and 7\,kpc, though we prefer 4.5\,kpc due to the slightly better alignment with the `radio-quiet' branch objects. However, we are unable to give a more robust determination of the distance from only three data points. The exact date of transition back to the `hard' state is not clear, which could help constrain the distance. The paucity of data points for EXO 1846$-$031 on the radio:X-ray plane shows further the need for rapid followup of X-ray binaries when they go into outburst and better monitoring during the hard state decay back to quiescence to better characterise this diagram.

\subsection{Energy from radio flares}
\label{sec:flares}

In Section~\ref{sec:radiodisc}, we identified two times when rapid increases in radio flux were observed in the AMI-LA data: flare 1 from MJD 58701.841--58702.876 where the flux increased from $5.9\pm0.5$ to $17.4\pm1.4$\,mJy beam$^{-1}$, and flare 2 from MJD 58724.811--58730.811, where the flux increased from $4.7\pm0.4$ to $22.9\pm1.8$\,mJy beam$^{-1}$ (see Fig.\ref{fig:lc_appendix}). It is possible to calculate the energy and power imparted into this radio flare by assuming equipartition between magnetic fields and particles \citep[][]{Longair,Fender2006}. In this case, the emission from the protons in the plasma is assumed negligible and we constrain the volume of the emitting component using the rise time of the flare e.g. equation 9.1--9.5 in \citet{Fender2006}. We use the AMI-LA central frequency of 15.5\,GHz and a distance of 4.5\,kpc to estimate the energies and magnetic fields.

For the first radio flare, we calculate $E_{\rm min} \sim$ ~ 4$\times$10$^{41}$ erg and hence a $P_{\rm min} \sim$ 5$\times$10$^{36}$ erg s$^{-1}$. This power is $\sim$10 per cent of the X-ray luminosity on the same day and is similar to the finding for the BHXB MAXI J1348$-$630 \citep[][]{Carotenuto2021b}. The calculated magnetic field corresponding to the minimum energy is $B_{\rm min} \sim$ 14 mG, with Lorentz factor $\gamma_{\rm e} \sim$ 1000. These values are similar to those found for MAXI J1348$-$630, ($B_{\rm min} \sim$ 10 mG, $\gamma_{\rm e} \sim$ 5000, \citealt[][]{Carotenuto2021b}), though in between the values calculated for the BHXB MAXI J1820+070 \citep{Bright2020,Espinasse2020} ($B_{\rm min} \sim$ 0.2 mG, $\gamma_{\rm e} \sim$ 5000) and Cyg-X3 ($B_{\rm min} \sim$ 500 mG, $\gamma_{\rm e} \sim$ 150, \citealt{Fender2006}), placing our values in a reasonable range compared to other BHXB flares. 
%though larger than the values calculated for the BHXB MAXI J1820+070 \citep{Bright2020}. 
The same calculations for the second flare yield an order of magnitude larger minimum energy: $E_{\rm min} \sim$ ~ 5$\times$10$^{42}$ erg, but similar power due to the longer rise time: $P_{\rm min} \sim$ 9$\times$10$^{36}$ erg s$^{-1}$. The corresponding magnetic field is $B_{\rm min} \sim$ 3 mG and Lorentz factor is $\gamma_{\rm e} \sim$ 2200.

\subsection{The presence of discrete ejecta and XB system parameters}
\label{sec:JetSpeed}

In Section~\ref{sec:radiodisc}, we showed that the third and sixth VLA panels in Fig.~\ref{fig:radioejecta} display clear signatures of jet ejecta, which are not detected in the fourth and fifth panels. We now discuss these epochs in turn to understand the cause of these components. In order to calculate jet component speeds, we first define an origin position using the first VLA observation, when EXO 1846$-$031 was in the `hard' state and an unresolved point source. This position is labelled with a white `plus' symbol in Fig.~\ref{fig:radioejecta}. 

The radio source in epoch three of Fig.~\ref{fig:radioejecta} is spectrally-steep ($\alpha = -1.5 \pm 0.3$), spatially-decoupled eastwards from the origin by 0.13-arcsec and it likely corresponds to the approaching jet ejected from the BH, though the uncertainty in this measurement is 0.05-arcsec (see Section~\ref{sec:radiodisc}) and we re-iterate that this component is offset by 2.5$\sigma$ from the nominal BH position. If we assume that the jet ejection happened around the transition from the `hard' state to the IMS (MJD 58700), this implies a movement of $\sim$15\,mas/day. 
Radio jet ejections can occasionally be resolved during the hard-to-soft transition \citep[e.g.][]{FenderBelloniGallo, Fender2009}, so the detection of component movement is unsurprising. As we do not know the exact jet launch date, we do not attempt to put further constraints on the jet speed at this point.

The jet component in epoch three is not detected in the following two epochs, 23 and 46 days respectively after the transition to the IMS and when the BH was in the `soft' state. Radio emission in BHXBs during the `soft' state is not often observed, as the steady jet emission is quenched by several orders of magnitude \citep[][\citealt{Russell2019}]{Coriat2011,Russell2011}. However, unresolved radio emission can be detected near the binary position in the case of long-lived jet ejecta, which would not be resolved in low resolution radio images, \citep[e.g.][]{Bright2020}. Assuming the jet continued to move at the same speed since epoch three, we would expect to detect a source  $\sim$0.34-arcsec and $\sim$0.65-arcsec, respectively, offset to the east from the origin in epochs four and five (red crosses in Fig.~\ref{fig:radioejecta}). The slight eastward extension in epoch four shows that we may have detected the faint remnants of the jet which has faded rapidly. We made further images with different image weightings in order to increase the significance of such possible jet structure, with no success. Similarly, we attempted a two component fit but were unable to obtain suitable fit results.

While the radio component detected in epoch five is coincident with the origin, it is resolved in the east-west direction, with a source size of 246\,mas $\times$ 123\,mas and position angle of 99$^{\circ}$. As EXO 1846$-$031 was clearly in the X-ray `soft' state during epochs four and five, we suggest that these epochs show faint jet emission that has been ejected from the BH, but cannot be resolved from the origin owing to the angular resolution of the VLA data. It is plausible that further jet ejections may have occurred during the soft state, possibly after the second radio flare around MJD 58730, which confuses the localisation of individual jet components. Higher resolution data would be required to resolve these components and test for multiple ejections. 

The sixth VLA epoch shows two components denoted by black circles in Fig.~\ref{fig:radioejecta}, the brighter of which is consistent with, but spatially-offset from the core position westwards by 0.04-arcsec, and a fainter component 0.85-arcsec eastwards of the core position. This eastern component is most likely the re-brightening jet component first observed in the third VLA epoch, having moved further away from the core. A re-brightening may occur if the jet interacts with a denser region of the ISM. If the eastward component travelled at the same speed between the third and sixth epochs, then this corresponds to $\sim$11\,mas day$^{-1}$. This value is similar to the initial $\sim$15\,mas$^{-1}$ observed movement estimated in epoch three, suggesting that these are the same components observed at different times. A linear fit to the separation of the eastern jet component over time would require an ejection date of $\sim$MJD 58697, when the source was clearly still in the `hard' state as evidenced by the X-ray data, so it is possible that some deceleration may have occurred during the propagation of this jet component, which is consistent with the red cross position in Fig.~\ref{fig:radioejecta}. 

The VLA data lack a clear detection of a receding jet. However, we can put limits on where we would expect it to reside by making assumptions on the distance and jet inclination. We assume a distance of 4.5\,kpc, given the discussion in Section~\ref{sec:distance}. Two inclination angles are suggested in the literature: \citep{Draghis2020} propose a high inclination of 73$^{\circ}$ whereas \citep{Wang2021} argue for a lower inclination of 40$^{\circ}$, from spectral fits of the same \textit{NuSTAR} data. Empirically, `radio-quiet' objects are found to have high-inclinations \citep[][]{Motta2018}, so we use the value of 73$^{\circ}$. Given the above assumptions and from the equations in \citet{MirabelRodriguez94,Fender2006}, the intrinsic speed of the approaching jet would be $\beta_{\rm int} \approx 0.29c$. Assuming that the receding jet was launched at the same speed in the opposite direction, it should be located 700--750\,mas away from the BH\footnote{We note that for a different combination of assumptions, e.g. $\theta$=40$^{\circ}$ and $d = 7.0$~kpc, we would have still have seen a receding component $>$400\,mas away from the original position in our final VLA epoch, which we clearly do not observe.}. As no component is detected in our VLA data at this distance, we argue that the surface brightness of the receding jet might be below our detection threshold. Using equation 2 in \citet{MirabelRodriguez94} to estimate the flux of the receding component in our last VLA epoch\footnote{We use a value of $k = 3$ for a discrete condensation in this calculation, and the spectral index and approaching jet flux from Table~\ref{tab:spec_index} and Table~\ref{tab:radiodataMeerKAT}, respectively.}, we find that the source brightness would be of the order of our image r.m.s. noise level. Therefore we would not have expected to detect this receding component at this late time. Thus, we suggest that the westward component in epoch six (MJD 58776) might be caused by secondary jet ejecta from a receding jet, but are unable to perform any further analysis.

\section{Conclusions}

We presented the radio and X-ray evolution of the X-ray binary (XB) EXO 1846$-$031 during its 2019 outburst, based on quasi-simultaneous X-ray (\textit{Swift} and \textit{MAXI}) and radio (1.3\,GHz MeerKAT, 4--8\,GHz VLA and 15.5\,GHz AMI-LA) observations. As part of the ThunderKAT X-ray binary monitoring program we observed EXO 1846$-$031 during 23 weekly-spaced epochs, and we obtained 19 radio detections. By studying the radio and X-ray light curves, the Hardness-Intensity Diagram (HID) and radio:X-ray plane, we showed that the evolution of EXO 1846$-$031 during its 2019 outburst was broadly consistent with typical `radio-quiet' black hole X-ray binary outbursts. 

The outburst began around MJD 58685 and proceeded to undertake a full hysteresis of the HID, transitioning from the X-ray spectral `hard' state on approximately MJD 58700. A radio flare was observed and peaked on MJD 58702 before the source transitioned to the X-ray `soft' state around MJD 58718. Following a decrease in the radio and X-ray light curves, a secondary radio flare was observed and peaked at approximately MJD 58730. After several months in the `soft' state, EXO 1846$-$031 transitioned back into the `hard' state some time before MJD 58906 at low X-ray flux levels, but the exact date of transition was not covered by \textit{Swift/XRT}.
%We studied the radio and X-ray light curves and Hardness-Intensity Diagram (HID) for the 2019 outburst, which began around MJD 58685 and completed a full hysteresis of the HID. 
%The source rose in X-ray flux while in the X-ray spectral `hard' state until approximately MJD 58700, upon which it transitioned into the `hard-intermediate' state for 2.5 weeks. On MJD 58701--58702, a radio flare in the AMI-LA data was observed, increasing from $5.9\pm0.5$ to $17.4\pm1.4$\,mJy beam$^{-1}$. The X-ray and radio fluxes then decreased before moving to the `soft' state around MJD 58718. During the `soft' state, the X-rays reached a secondary peak around MJD 58736. This secondary X-ray peak was mirrored in the radio, with a second AMI-LA radio flare observed from MJD 58724.811--58730.811, where the flux increased from $4.7\pm0.4$ to $22.9\pm1.8$\,mJy beam$^{-1}$. After several months in the `soft' state, EXO 1846$-$031 transitioned back into the `hard' state some time before MJD 58906 at low X-ray flux levels, but the exact date of transition was not covered by \textit{Swift/XRT} monitoring and AMI-LA radio observations did not detect EXO 1846$-$031 around this time.

Since a precise distance measurement of EXO 1846$-$031 was not available, we first calculated a distance from the soft-to-hard state transition luminosity distribution of BHXBs of 2.4--7.5\,kpc \citep{Kalemci2013}. We attempted to estimate the distance from the available X-ray `hard' state radio and X-ray data using the radio:X-ray plane. Using this method, we showed that the gradient of the radio:X-ray correlation for EXO 1846$-$031 was $\sim$1.0 but were unable to constrain the distance estimate further; the data align best with a distance of $\sim$4.5\,kpc, but the paucity of data points prevents any meaningful uncertainties. A BH primary is preferred to a NS in EXO 1846$-$031, as evidenced by the distance \citep[][]{Miller-JonesATel} and the trends on the radio:X-ray plane: if EXO 1846$-$031 is a NS, it would be one of the radio-brightest NSs to be discovered.

%Throughout the outburst, the radio and X-ray light curves appear mostly correlated. 
We estimated equipartition minimum energies of each of the observed AMI-LA radio flares to be $E_{\rm min} \sim$ 4$\times$10$^{41}$ and 5$\times$10$^{42}$ erg for each flare, respectively. These values and their associated powers ($P_{\rm min} \sim$ 5$\times$10$^{36}$ and 9$\times$10$^{36}$ erg, respectively) are broadly consistent with other radio jetted BHXBs in the literature, including MAXI J1348$-$630 \citep[][]{Carotenuto2021b} and MAXI J1820+070 \citep{Bright2020}.

We observed steepening radio spectral indices around MJD 58750 which suggest optically-thin radio emission, possibly from unresolved radio ejecta in all three radio light curves. The VLA observations detected the presence of an eastward jet component 0.85-arcsec from the core position. We thus calculate an intrinsic jet speed of $\beta_{\rm int} = 0.29c$ for the eastward component, assuming a distance of 4.5\,kpc and an inclination angle of $\theta$=73$^{\circ}$, but we were unable to conclusively detect a receding jet component, and are unable to put further strong constraints on the system geometry.  %We showed that our radio observations were not sensitive enough to detect a receding jet given our observation times. 
We suggest that the observed radio components that are consistent with the XB core position while in the soft state are from unresolved, potentially additional jet ejections; higher resolution interferometers would have been required in this instance to conclusively resolve the receding jet emission.

The ThunderKAT X-ray binary monitoring programme has been operational for three years and is continuing to detect radio jet emission and discrete radio jet ejecta from BHXB systems. Historical systems like EXO 1846$-$031, where previous radio detections were not possible, are potential targets for ThunderKAT and offer the chance to further understand XB jet emission in a co-ordinated framework -- this work has highlighted the need for regular radio monitoring at high resolution to resolve jet emission and extract physical system parameters.

\section*{Acknowledgements}

This work was supported by the Oxford Centre for Astrophysical Surveys, which is funded through generous support from the Hintze Family Charitable Foundation. 
JvdE is supported by a Lee Hysan Junior Research Fellowship awarded by St. Hilda's College, Oxford. 
We thank the staff at the South African Radio Astronomy Observatory (SARAO) for their rapid scheduling of these observations. The MeerKAT telescope is operated by SARAO, which is a facility of the National Research Foundation, an agency of the Department of Science and Technology. We also acknowledge the staff who operate and run the AMI-LA telescope at Lord's Bridge, Cambridge, for the AMI-LA radio data. AMI is supported by the Universities of Cambridge and Oxford, and
acknowledges support from the European Research Council under grant ERC-2012-StG-307215 LODESTONE.
The National Radio Astronomy Observatory is a facility of the National Science Foundation operated under cooperative agreement by Associated Universities, Inc.
%\textit{e}-MERLIN is a National Facility operated by the University of Manchester at Jodrell Bank Observatory on behalf of STFC, part of UK Research and Innovation.
We acknowledge the use of public data from the \textit{Swift} data archive. This research has made use of MAXI data provided by RIKEN, JAXA and the MAXI team.
This research made use of APLpy, an open-source plotting package for Python \citep[][]{APLpy}. 
DRAW would like to acknowledge the help of Mark Lacy at the NRAO Help Desk for suggestions that greatly improved the VLA data calibration and imaging quality.

\vspace{-0.2cm}
\section*{Data Availability}

All of the X-ray data presented here (\textit{Swift/XRT}, \textit{Swift/BAT} and \textit{MAXI/GSC}) can be downloaded from the public archives, note in the manuscript. The radio data (MeerKAT and AMI-LA) can be made available on request to the corresponding author. The VLA data is publicly available on the NRAO VLA archive.



%\bibliographystyle{mnras}

\bibliographystyle{mnras}
\bibliography{bib}

%%%%%%%%%%%%%%%%%%%%%%%%%%%%%%%%%%%%%%%%%%%%%%%%%%

%%%%%%%%%%%%%%%%%%%%%%%%%%%%%%%%%%%%%%%%%%%%%%%%%%

%%%%%%%%%%%%%%%%% APPENDICES %%%%%%%%%%%%%%%%%%%%%

\appendix

\section{Zoom plot of the radio and X-ray light curves}

\begin{figure*}
	\includegraphics[width=0.95\textwidth]{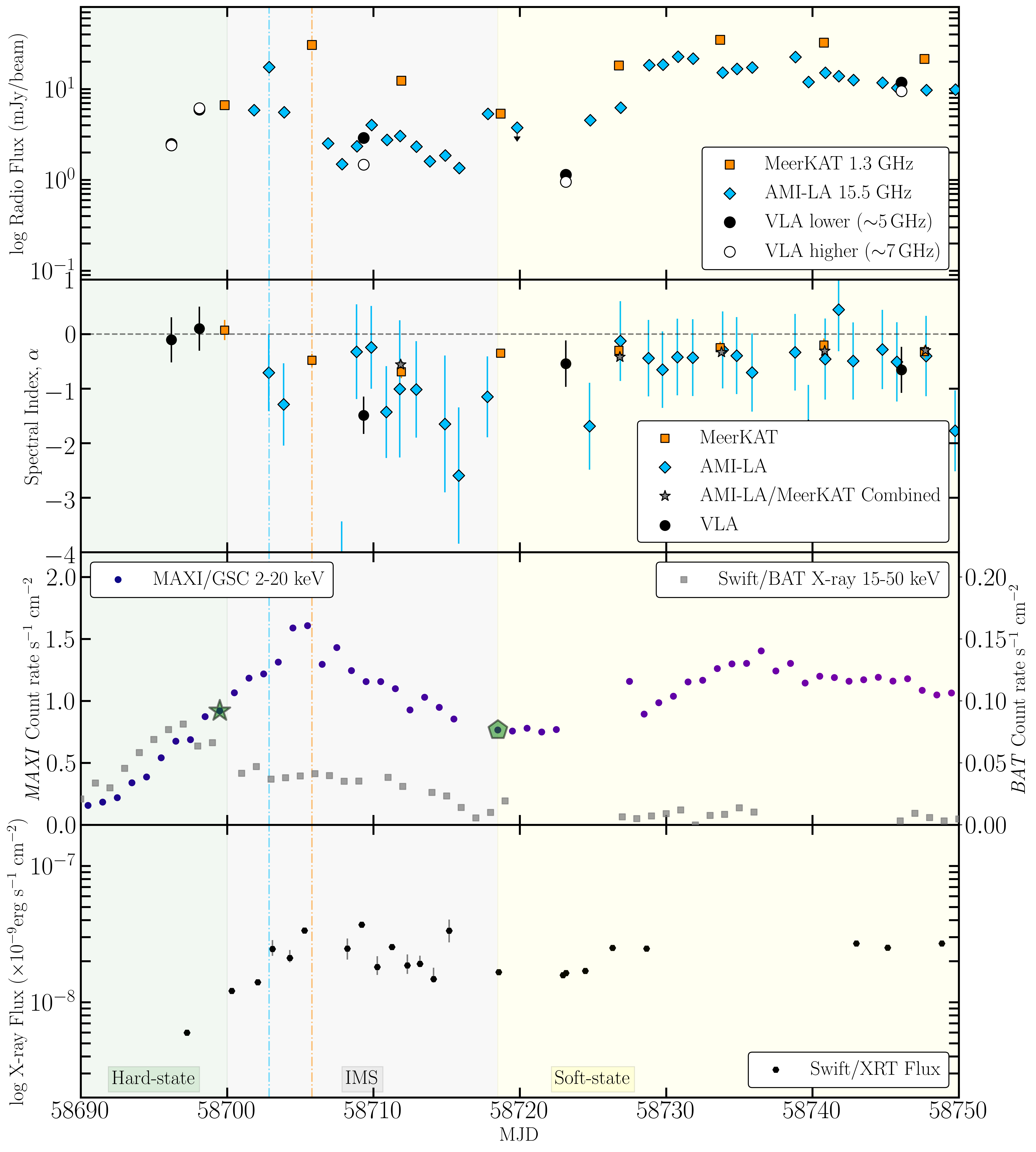}
    \caption{
    Radio and X-ray light curves and spectra of the 2019 outburst of the BHXB EXO 1846$-$031, for the region between MJD 58690 and MJD 58750. The panels are identical to Figure~\ref{fig:lc} with minor changes to the $y$ axes to better represent the data: the top three panels have an identical y axis scale to Figure~\ref{fig:lc}, but the bottom panel (\textit{Swift/XRT} fluxes) are restricted to the range between 10$^{-7}$ -- 10$^{-9}$ erg, where Figure~\ref{fig:lc} extended to 10$^{-11}$ erg. %The top panel shows the radio light curves for AMI-LA (blue diamonds), MeerKAT (orange squares) and VLA in the lower band near 5\,GHz (black circles) and upper band near 7\,GHz (black unfilled circles). Upper limits are denoted by downward facing arrows. The peaks in the AMI-LA and MeerKAT light curves are shown by blue or orange dashed lines in all panels, respectively. The second panel shows the in-band spectral indices for AMI-LA and MeerKAT in the same colour/marker style, with grey stars denoting the AMI-LA/MeerKAT combined spectral index and the two-band VLA spectral index denoted by black circles. In both of the top two plots, the final VLA observation shows two sources (see Table~\ref{tab:VLAdata}), and we include both of these sources on this plot.
    %The bottom two panels show the X-ray data in order: the X-ray light curve of the outburst using all-sky monitor data from \textit{MAXI/GSC} and \textit{Swift/BAT}; the \textit{Swift/XRT} unabsorbed flux obtained from the X-ray spectral fits to the data. 
    %For the all-sky monitor data, the grey squares show the X-ray count rate from \textit{Swift/BAT} in the 15--50\,keV `hard' band, while the coloured circles show the X-ray count rate in the \textit{MAXI/GSC} 2--20\,keV `soft' band. The colour for these circles denotes the day of observing, for direct comparison to Fig.~\ref{fig:HID}. 
    %We represent the days where we consider the `hard' state transition as a green star in the fourth panel, the first point where the source is in the `soft' state as a green pentagon, and the green hexagon represents the first point where EXO 1846$-$031 is definitely back in the `hard' state: see discussion in Section 3.1. We label these the X-ray states in the bottom panel and shade the background green, grey or yellow for each of the X-ray states, respectively. We show with a black dashed line when EXO 1846$-$031 was Sun constrained for the \textit{Swift/XRT}.
    }
    \label{fig:lc_appendix}
\end{figure*}

%%%%%%%%%%%%%%%%%%%%%%%%%%%%%%%%%%%%%%%%%%%%%%%%%%

% Don't change these lines
\bsp	% typesetting comment
\label{lastpage}
\end{document}